\documentclass[letterpaper]{article} 
\usepackage{aaai25}  
\usepackage{times}  
\usepackage{helvet}  
\usepackage{courier}  
\usepackage[hyphens]{url}  
\usepackage{graphicx} 
\usepackage{natbib}  
\usepackage{caption} 
\usepackage{algorithm}
\usepackage{algorithmic}

\usepackage{newfloat}
\usepackage{listings}

\usepackage{multirow} 
\usepackage{makecell} 
\usepackage{bm} 
\usepackage{amsmath} 
\usepackage{amssymb} 

\DeclareCaptionStyle{ruled}{labelfont=normalfont,labelsep=colon,strut=off} 
\floatstyle{ruled}
\newfloat{listing}{tb}{lst}{}
\floatname{listing}{Listing}
\title{Few-Shot Domain Adaptation for Learned Image Compression}
\author{
    Tianyu Zhang,
    Haotian Zhang,
    Yuqi Li,
    Li Li,
    Dong Liu\thanks{Corresponding Author}
}
\affiliations{
    University of Science and Technology of China\\
    \{zhangtianyu,zhanghaotian,lyq010303\}@mail.ustc.edu.cn, \{lil1,dongeliu\}@ustc.edu.cn
}

\usepackage{bibentry}

\begin{document}

\maketitle

\begin{abstract}
Learned image compression (LIC) has achieved state-of-the-art rate-distortion performance, deemed promising for next-generation image compression techniques. However, pre-trained LIC models usually suffer from significant performance degradation when applied to out-of-training-domain images, implying their poor generalization capabilities. To tackle this problem, we propose a few-shot domain adaptation method for LIC by integrating plug-and-play adapters into pre-trained models. Drawing inspiration from the analogy between latent channels and frequency components, we examine domain gaps in LIC and observe that out-of-training-domain images disrupt pre-trained channel-wise decomposition. Consequently, we introduce a method for channel-wise re-allocation using convolution-based adapters and low-rank adapters, which are lightweight and compatible to mainstream LIC schemes. Extensive experiments across multiple domains and multiple representative LIC schemes demonstrate that our method significantly enhances pre-trained models, achieving comparable performance to H.266/VVC intra coding with merely 25 target-domain samples. Additionally, our method matches the performance of full-model finetune while transmitting fewer than $2\%$ of the parameters.
\end{abstract}

%

\section{Introduction}
With the rapid growth of images in multimedia, image compression has become essential for efficient storage and transmission. Over the past decade, conventional methods including JPEG \cite{wallace1991jpeg}, H.265/HEVC \cite{sullivan2012overview} and H.266/VVC \cite{bross2021overview}, have significantly contributed to the real-world applications of image compression. Recently, learned image compression (LIC) has achieved impressive progress in rate-distortion (RD) performance using non-linear transforms and end-to-end optimization. Several studies \cite{he2022elic, jiang2023mlicpp} have even surpassed VTM (the reference software for H.266/VVC) on both PSNR and MS-SSIM, indicating LIC's potential for future image compression techniques.

\begin{figure}[t]
  \includegraphics[width=0.48\textwidth]{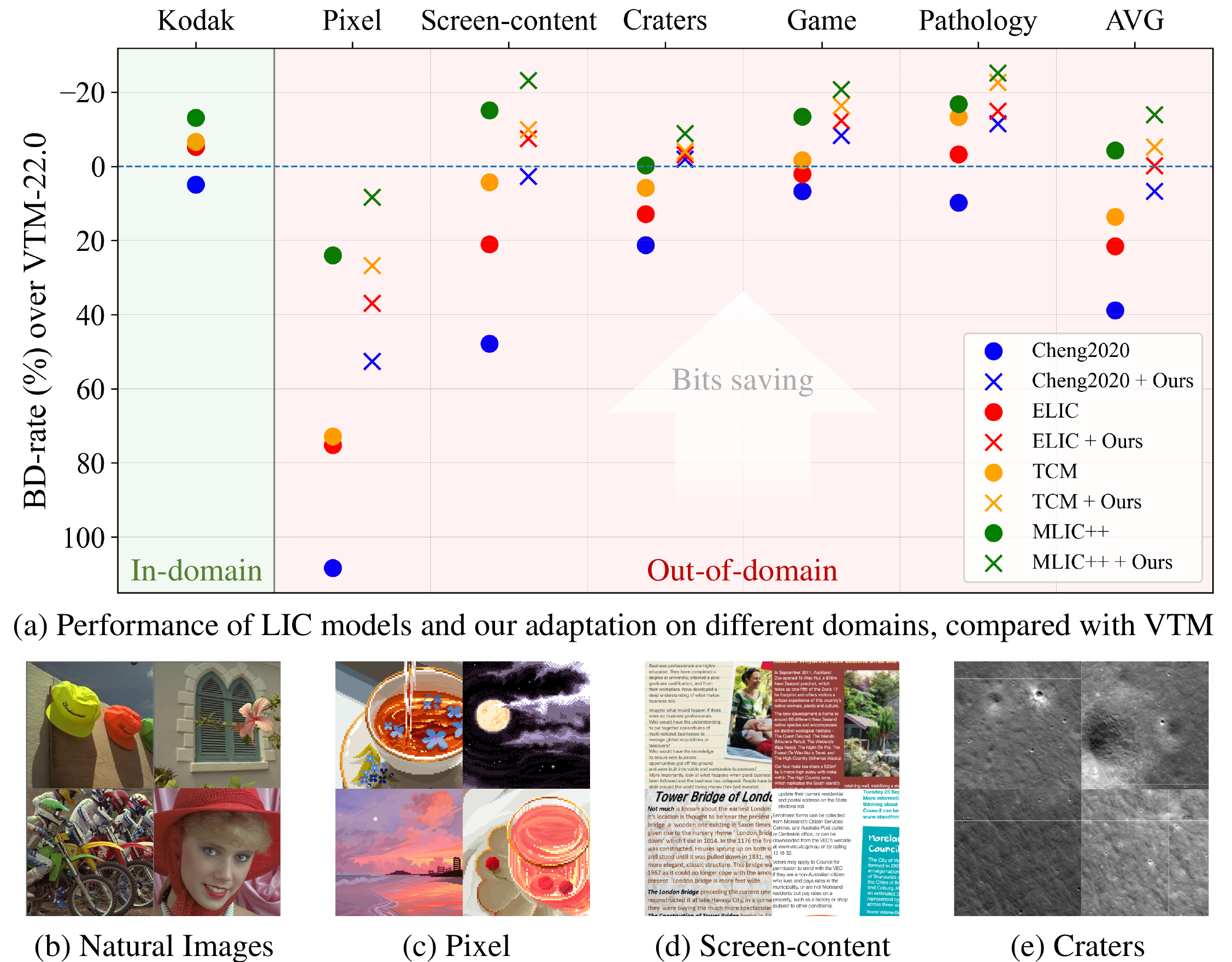}
  \caption{(a) BD-rate ($\downarrow$) of four advanced LIC models with or without our method on different domains. (b)-(e) Images from different domains have visible different characteristics.}
  \label{fig1}
\end{figure}

Despite the encouraging progress, some studies \cite{shen2023dec, lv2023dynamic} have pointed out the performance degradation of pre-trained LIC models on out-of-domain images (we simplify in/out-of-training-domain as in/out-of-domain for convenience). It is common practice to train and evaluate LIC models on natural image benchmarks \cite{kodak, agustsson2017ntire, asuni2014testimages, clic, li2024ustc}. However, practical codecs may well be challenged by images from various domains, many of which have significant domain gaps compared to natural images. For demonstration purposes, we test advanced LIC models including Cheng2020 \cite{cheng2020learned}, ELIC \cite{he2022elic}, TCM \cite{liu2023learned}, and MLIC++ \cite{jiang2023mlicpp} in Fig. \ref{fig1}. While all these models perform comparably or better than the conventional codec VTM on natural images, they exhibit varying degrees of performance drop and generally fall behind VTM on out-of-domain images such as pixel-style art and screen content, highlighting the necessity for efficient adaptation methods for pre-trained LIC models.

In this regard, most existing methods \cite{shen2023dec, lv2023dynamic, tsubota2023universal} suggest instance adaptation (IA), which aims to adapt LIC models to a single image at inference. Despite its effectiveness, IA does not perform actual domain adaptation (DA) for pre-trained models, and is time-consuming and computationally expensive due to its per-image online training. On the other hand, \citet{katakol2021danice} proposed DA for LIC using limited target samples. They suggested selective finetune, and adapted pre-trained models to a target domain with one-time training and transmission, demonstrating great adaptation efficiency for deployed models. Nevertheless, the proposed method is restricted to GDN structures \cite{DBLP:journals/corr/BalleLS15}, and suffers from large additional parameters on recent schemes \cite{he2022elic, liu2023learned, jiang2023mlic}.

Despite the promising applications like codec calibration for batch transmission, DA for LIC remains largely unexplored. Recently, \citet{presta2024domain} pre-trained a gate network on multiple domains, enhancing the generalization capacity of LIC models. However, they did not address the problem of adapting LIC models to a specific domain. Therefore, in this paper, we focus on \textbf{adapting pre-trained LIC models to a target domain}. For practicality, we follow \citet{katakol2021danice} and study DA for LIC with limited samples. We also highlight the versatility of the DA method to different domains and mainstream LIC schemes.

Building on these insights, we propose a universal few-shot domain adaptation method for LIC using compact adapters, achieving superior performance across various domains and mainstream LIC schemes. Inspired by \cite{li2023revisiting}, we explore domain gaps in LIC through disturbed channel-wise decomposition. We demonstrate that pre-trained LIC models suffer from scattered channel-wise energy allocation on out-of-domain images. To address this problem, we introduce convolution-based adapters (Conv-Adapters) and low-rank adapters (LoRA-Adapters) to the pre-trained models for channel-wise re-allocation. These adapters are trained with a few target samples, and are plug-and-play for processing specific domains.

Our main contributions can be summarized as follows:
\begin{itemize}
\item We propose a universal few-shot domain adaptation method for learned image compression by incorporating compact adapters. Our method is applicable to different domains and mainstream LIC schemes.
\item We explore the domain gaps in LIC through disturbed channel-wise decomposition, which results in scattered energy allocation on out-of-domain images. The proposed channel-wise re-allocation with adapters strengthens energy compaction.
\item Extensive experiments verify the effectiveness of our method across mainstream domains and LIC schemes. We generally enhance pre-trained models to VTM on every domain, achieving comparable improvements to full-model finetune with tiny additional parameters.
\end{itemize}

\section{Related Work}

\subsection{Learned Image Compression}

    Learned image compression has demonstrated competitive potential against traditional codecs. \citet{DBLP:conf/iclr/BalleLS17} proposed the first end-to-end LIC model with non-linear transforms and uniform quantizer, which was further strengthened by hyperprior \cite{DBLP:conf/iclr/BalleMSHJ18} and context model \cite{minnen2018joint}. \citet{cheng2020learned} first achieved comparable performance with VTM using Gaussian mixture likelihoods. \citet{he2022elic} proposed uneven channel autoregressive model, demonstrating pleasant performance and complexity. Recently, \citet{liu2023learned} designed Transformer-CNN mixture block to incorporate the ability of both structures for LIC. \citet{jiang2023mlic} introduced multi-reference context to capture different correlations, achieving the state-of-the-art performance \cite{jiang2023mlicpp}.

\begin{figure}[tbp]
  \includegraphics[width=0.48\textwidth]{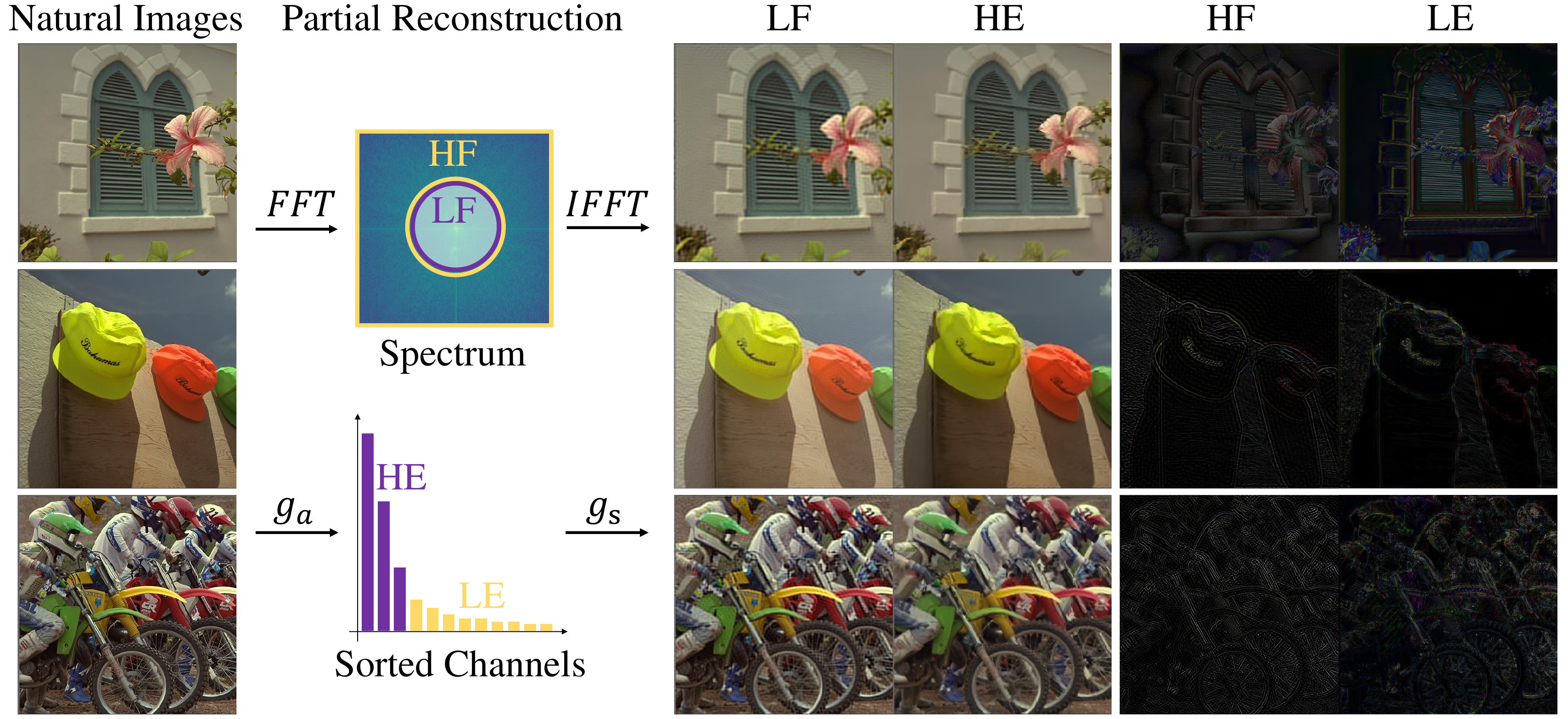}
  \caption{Analogy between frequency components and channels of LIC latents. For one image, we perform Fast Fourier Transform (FFT) and then reconstruct from low-frequency (LF) or high-frequency (HF) components respectively. Similarly, we perform a learned analysis transform ($g_a$) and then reconstruct from high-energy (HE) or low-energy (LE) channels respectively. Please view on screen and zoom in to observe the reconstructions from HF/LE.} %
  \label{relation}
\end{figure}

In the regard of interpretability, early work \cite{DBLP:conf/iclr/BalleMSHJ18, cheng2020learned} visualized the latent in LIC to evaluate the preciseness of the entropy model. \citet{he2022elic} sorted latent channels by energy, suggesting low-frequency components concentrated on a few channels while the rest was extremely sparse. \citet{duan2022opening} interpreted latent channels as orthogonal transform coefficients, while \citet{li2023revisiting} suggested frequency decomposition is an intrinsic property for LIC. Nevertheless, these results on latent representation did not fully consider out-of-domain images.

\begin{figure*}[!tbp]
\centering
  \includegraphics[width=0.9\textwidth]{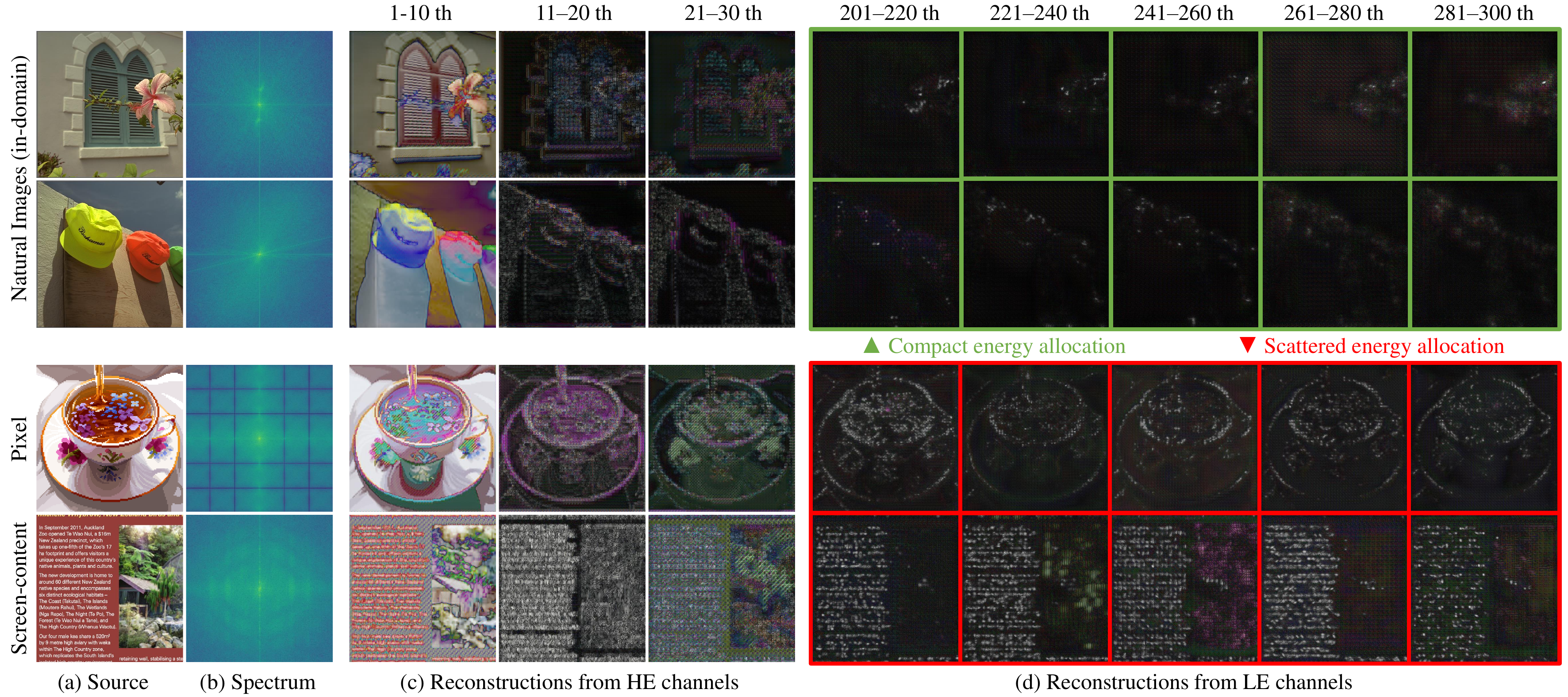}
  \caption{Following Fig. \ref{relation}, we explore the domain gaps by observing the channel-wise decomposition. The top two rows display in-domain natural images, while the bottom two rows show out-of-domain images. (a) Source image. (b) Spectrum (using FFT). (c, d) Reconstructions from different HE channels and LE channels respectively. Here the total number of channels is 320. \emph{Out-of-domain images have more HF components as shown in (b), as well as more information embedded into LE channels as shown in (d). Thus, our key idea is to re-allocate the information from LE channels to HE channels for out-of-domain images.}}
  \label{fig2}
\end{figure*}

\subsection{Adaptation for Learned Image Compression}

Two kinds of adaptation namely instance adaptation (IA) and domain adaptation (DA) have been studied for LIC. IA \cite{DBLP:conf/cvpr/CamposMDS19, vanrozendaal2021instanceadaptive, tsubota2023universal, shen2023dec, lv2023dynamic} adapts pre-trained models to a single image rather than a domain using per-image online training. Despite its effectiveness, IA is computationally expensive and time-consuming as shown in \textbf{Appendix A}. Contrarily, \citet{katakol2021danice} introduced the problem of DA for LIC, adapting pre-trained LIC models to a target domain with one-time training. They suggested finetuning GDN layers with channel-wise parameters and the entire entropy model, which was restricted by specific structures, and additional parameters due to the complex entropy models. We alleviate this issue by designing compact and universal adapters. 

\section{Method}

\subsection{Motivation}

Recent studies \cite{duan2022opening, li2023revisiting} suggest that the transform in LIC can be interpreted as channel-wise frequency decomposition in a non-linear fashion. We compare latent channels to frequency components of a source image using linear transforms (e.g. FFT), as reconstructions from partial channels or partial frequency components demonstrate obvious similarities in Fig. \ref{relation}. Motivated by this, we further study the channel-wise decomposition on different domains by reconstructions from specific channels in Fig. \ref{fig2} (c) and (d). Consistent with previous findings \cite{he2022elic, duan2022opening, li2023revisiting}, decomposed information concentrates on high-energy channels (respectively low-frequency). For in-domain images, the energy allocation on channels is more compact, with little information of the source images found in low-energy channels. However, for out-of-domain images, more high-frequency components are presented in the spectrum (Fig. \ref{fig2} (b)), and obvious contours of source images can be observed in reconstructions from low-energy channels, as shown in Fig. \ref{fig2} (d). This indicates a scattered energy allocation of the pre-trained LIC model on out-of-domain images, which differs from the compact energy allocation on in-domain images.

Given these channel-wise differences among latents of in-domain and out-of-domain images, we assume that the following entropy estimation and reconstruction modules can not operate effectively in the pre-trained manner due to improper latents, leading to degraded rate-distortion performance. Therefore, our key idea is to perform channel-wise re-allocation, which concentrates the information from LE channels to HE channels for out-of-domain images. Specifically, we propose gently re-allocating the channels of intermediate latents by inserting compact adapters into the pre-trained models. The implementation and performance of these adapters are detailed in the following sections.

\subsection{Channel-wise Re-allocation with Adapters}

\begin{figure*}[!htbp]
\centering
  \includegraphics[width=0.9\textwidth]{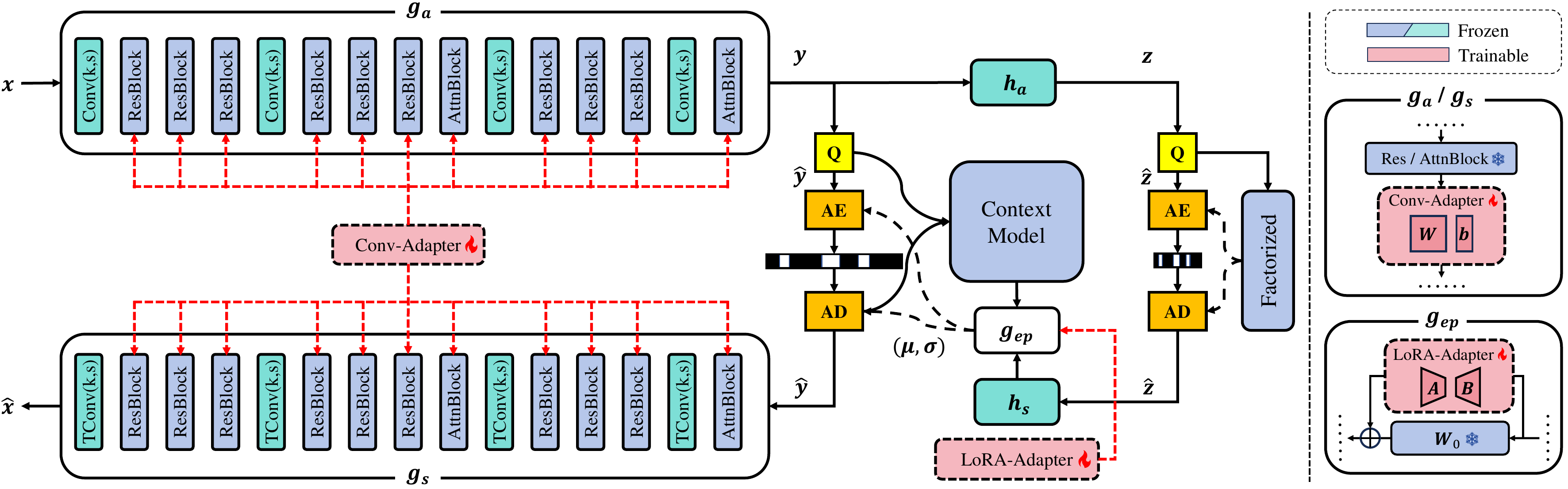}
  \caption{Deployment of our method on ELIC \cite{he2022elic}. We denote $g_{ep}$ as the entropy parameters network, while the other notations follow the same explanations in \cite{he2022elic}. As detailed on the right, Conv-Adapters are inserted serially after non-linear blocks in the transform, while LoRA-Adapters are added to the pre-trained weight matrix $W_{0}$ in $g_{ep}$. $W$ and $b$ are weight and bias of Conv-Adapter, respectively, while $A$ and $B$ are low-rank matrices. Only adapters are trainable.}
  \label{overview img}
\end{figure*}

We present the deployment of our method on ELIC \cite{he2022elic} in Fig. \ref{overview img}. Two types of adapters, namely Conv-Adapters and LoRA-Adapters, are introduced. To refine the channel-wise decomposition in the transform, we insert Conv-Adapters after non-linear blocks. For entropy estimation, LoRA-Adapters are applied to the entropy parameters network. We highlight the compatibility of the proposed method to mainstream LIC schemes. 

\subsubsection{Conv-Adapters}

We introduce Conv-Adapters to the transform to adjust the channels of intermediate latents. Denoting the output latent of a specific block as $L {\in} \mathbb{R}^{c \times h \times w}$ and its $ith$ channel $L_{i} {\in} \mathbb{R}^{h \times w}$, we aim to re-allocate the information on channels before passed to the next block:
\begin{equation}
  L^{\prime} = (L^{\prime}_{0}, L^{\prime}_{1}, ..., L^{\prime}_{c-1}) = f(L_{0}, L_{1}, ..., L_{c-1})
\end{equation}
where $L^{\prime}$ denotes the output latent with refined information allocation on channels, and $f$ denotes the re-allocation mechanism introduced by Conv-Adapters. Considering the restriction on computational complexity and limited samples, we simplify this mechanism into a channel-wise linear transform, and apply Conv1$\times$1 as the Conv-Adapter. To maintain the training stability, we initialize the weight matrix $W {\in} \mathbb{R}^{c \times c \times 1 \times 1}$ of Conv1$\times$1 to identity matrix $I {\in} \mathbb{R}^{c \times c \times 1 \times 1}$, and the bias $b {\in} \mathbb{R}^c$ to zeros. Therefore, we refine intermediate latents in the transform by the following linear re-allocation:
\begin{equation}
  L^{\prime} = W \cdot (L_{0}, L_{1}, ..., L_{c-1})^{T} + b
\end{equation}

\subsubsection{LoRA-Adapters}

To accommodate the channel-wise re-allocation in the transform, we adopt Low-Rank Adaptation (LoRA) \cite{DBLP:conf/iclr/HuSWALWWC22, lv2023dynamic} in the entropy model. Mainstream LIC schemes have large variances on the entropy model, resulting in difficulty for a universal adaptation method on entropy estimation. Besides, hidden dimensions in the entropy model are empirically large for better performances, indicating convolution-based adapters will boost the amount of additional parameters. Therefore, we introduce low-rank matrices (LoRA-Adapters) to the common entropy parameters network. Namely, we modify the weight of a pre-trained layer $W_{0} {\in} \mathbb{R}^{c_{out} \times c_{in} \times h \times w}$ by:
\begin{equation}
\begin{aligned}
    W^{\prime}_{0} &= W_{0} + \Delta W \\
    \Delta W &= BA
\end{aligned}
\end{equation}
where $A {\in} \mathbb{R}^{r \times c_{in}}$ is initialized with Gaussian, $B {\in} \mathbb{R}^{c_{out} \times r}$ is set to zeros, ensuring $\Delta W = 0$ thus the training stability. The rank $r \ll \min\left\{c_{in}, c_{out}\right\}$, and is set to $10$ empirically. LoRA-Adapters serve as bias on channel dimension for pre-trained weights, and can be merged once trained without introducing additional latency at inference.

\begin{figure}[tbp]
\centering
  \includegraphics[width=0.44\textwidth]{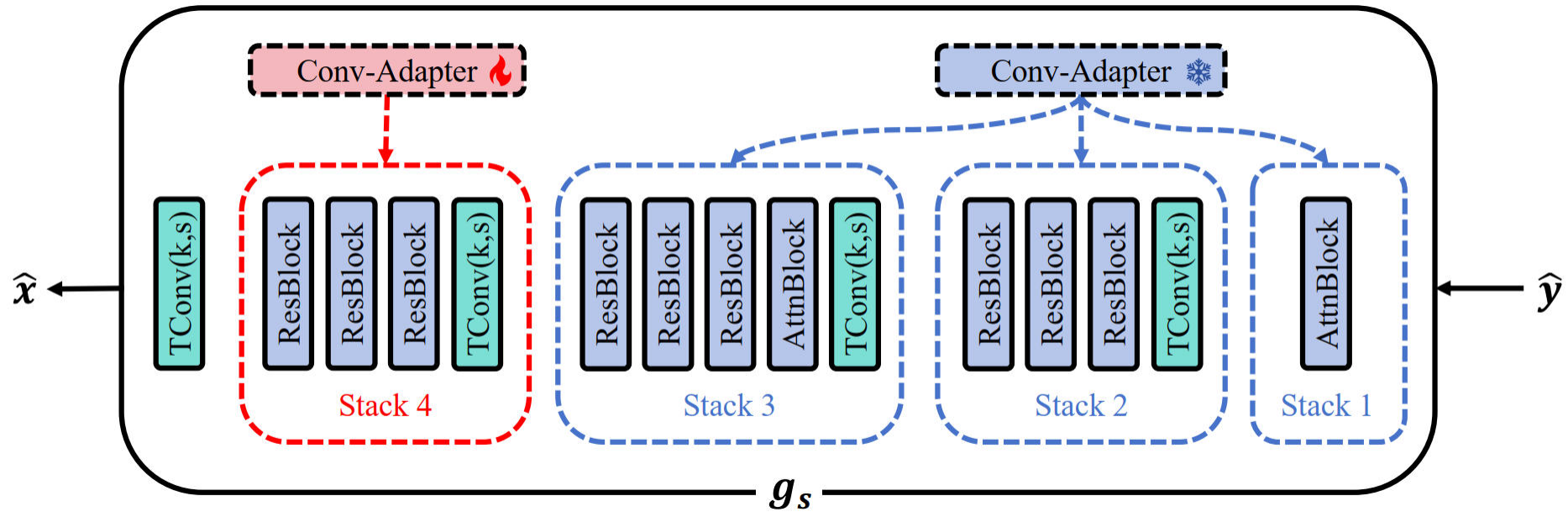}
  \caption{Division of $g_{s}$ into four stacks. In the second stage, we only finetune adapters in Stack 4 for reconstruction.}
  \label{stage2}
\end{figure}

\subsection{Two-Stage Training Strategy}

The training of adapters consists of two stages, as detailed in Fig. \ref{overview img} and Fig. \ref{stage2} respectively. We denote $\theta = \left\{W, b, A, B\right\}$ as the total parameters of adapters. Given the analysis transform $g_{a}$, synthesis transform $g_{s}$, hyper analysis transform $h_{a}$, uniform
scalar quantization $Q$, and $\lambda$ to balance bit-rate $\mathcal{R}$ and distortion $\mathcal{D}$, we first freeze all pre-trained parameters $\phi$ and train adapters $\theta$ jointly with:
\begin{equation}
  \mathcal{L}(\theta) = \mathcal{R}(\hat{y}) + \mathcal{R}(\hat{z}) + \lambda \cdot \mathcal{D}(x, \hat{x})
\end{equation}
where:
\begin{equation}
\begin{aligned}
  \hat{y}&=Q(g_{a}(x;\phi,\theta)) \\
  \hat{x}&=g_{s}(\hat{y};\phi,\theta) \\
  \hat{z}&=Q(h_{a}(y;\phi))
\end{aligned}
\end{equation}

We suggest adapters optimized by a joint objective function suffer from insufficient training under limited samples. Inspired by the two-stage training strategy in \cite{guo2021soft}, we further conduct a second training stage. Concretely, we divide $g_{s}$ into four stacks by upsample operations, as shown in Fig. \ref{stage2}. We freeze all the adapters except the ones in Stack 4, and finetune this small set (denoted as $\Delta\theta$) with hard quantization and distortion only:
\begin{equation}
  \mathcal{L}(\Delta\theta) = \mathcal{D}(x, \hat{x}), \text{switch $Q$ to $rounding$.}
\end{equation}

We demonstrate in Section 4.3 that this stage is simple but effective without introducing more parameters. Once trained, adapters in $g_{a}$ are stored at the encoder, while the rest should be transmitted to the decoder. In this way, a full set of adapters corresponds to a certain domain, and is plug-and-play for flexibly processing specific domains.

\begin{figure*}[!ht]
\centering
  \includegraphics[width=0.97\textwidth]{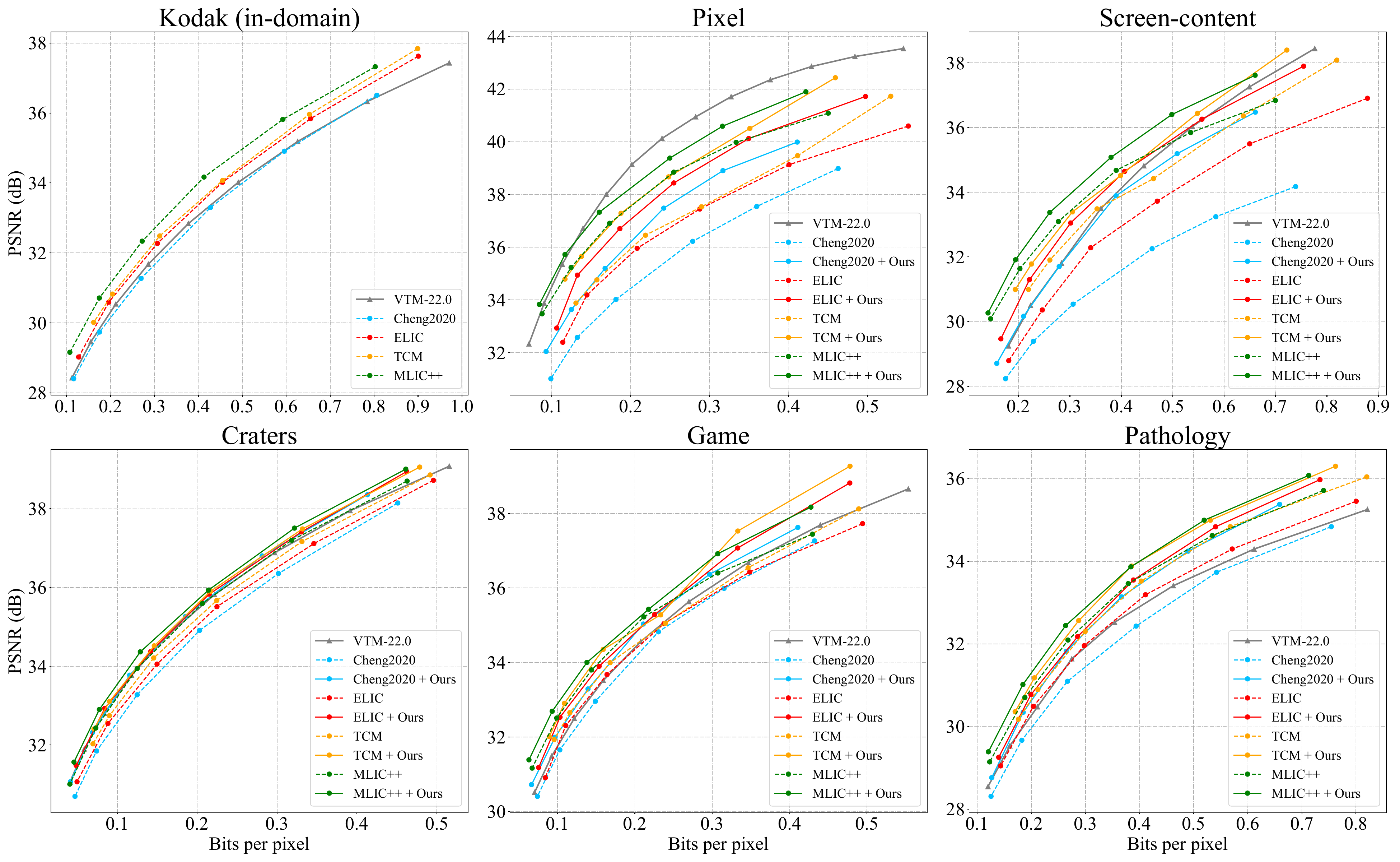}
  \caption{RD curves including VTM-22.0 intra coding, pre-trained LIC models, and models adapted with 25 samples in our method. We generally enhance pre-trained models to the level of VTM on the other five domains.}
  \label{rdcurve}
\end{figure*}

\section{Experiments}

\subsection{Settings}

\textbf{Domains and Datasets.} We follow \citet{lv2023dynamic, shen2023dec} and sort five domains with 150 images each, including (i) pixel-style images \cite{lv2023dynamic}; (ii) screen content images from SCID \cite{ni2017esim}, SCI1K \cite{yang2021implicit}, SIQAD \cite{yang2015perceptual}; (iii) craters images cropped from Lunar Reconnaissance Orbiter Camera (LROC) \cite{robinson2010lunar}; (iv) game images cropped from GamingVideoSET \cite{barman2018gamingvideoset}; (v) pathology images cropped from BRACS \cite{brancati2022bracs}. All crops have the same resolution of $600 \times 800$. We select a fixed test set with 100 images for each domain. We use Kodak \cite{kodak} for the natural image domain.

\textbf{Models.} We test our method on advanced LIC models, namely Cheng2020 \cite{cheng2020learned}, ELIC \cite{he2022elic}, MLIC++ \cite{jiang2023mlicpp} and TCM \cite{liu2023learned}. We use the official pre-trained weights except ELIC, for which we train ourselves for a lack of official weights. More details can be found in \textbf{Appendix F}.

\textbf{Implementation.} We train adapters with $N=\left\{5, 10, 25\right\}$  samples, respectively. We split the samples into training and validation set in a proportion of 4:1. We use a patch size of 256 and a batch size of 4. In the first stage, we set learning rate stages $\left\{50, 10, 7.5, 5, 2.5, 1\right\} \times 10^{-5}$. For different $N$, we further set max epoch $= \left\{250, 500, 750\right\}$. In the second stage, we train selected adapters for another max epoch with fixed learning rate $5 \times 10^{-4}$.

\begin{table}[t]
    \centering
    \setlength{\tabcolsep}{1mm}
    \begin{tabular}{ccccc}
        \hline
        Baseline & Params & Method & Transmit & Prop. \\
        \hline
        \multirow{3}{*}{Cheng2020} & \multirow{3}{*}{31.48M} & Finetune & 24.12M & 76.62\% \\
        & & DANICE & 10.86M & 34.50\% \\
        & & Ours & \textbf{0.37M} & \textbf{1.17\%} \\
        \hline
        \multirow{3}{*}{ELIC} & \multirow{3}{*}{40.72M} & Finetune & 30.98M & 76.08\% \\
        & & DANICE & / & / \\
        & & Ours & \textbf{0.71M} & \textbf{1.74\%} \\
        \hline
        \multirow{3}{*}{TCM} & \multirow{3}{*}{45.18M} & Finetune & 41.93M & 92.81\% \\
        & & DANICE & 35.17M & 77.84\% \\
        & & Ours & \textbf{0.88M} & \textbf{1.95\%} \\
        \hline
        \multirow{3}{*}{MLIC++} & \multirow{3}{*}{116.72M} & Finetune & 110.45M & 94.71\% \\
        & & DANICE & 98.41M & 84.31\% \\
        & & Ours & \textbf{0.73M} & \textbf{0.63\%} \\
        \hline
    \end{tabular}
    \caption{The number of additional parameters to transmit compared with full-model finetune and DANICE \cite{katakol2021danice}, Note that DANICE can not be applied to ELIC.}
    \label{tb1}
\end{table}

\subsection{Adaptation Performance}

\subsubsection{Comparisons with Traditional Codec}

We compare the proposed method with the conventional standard VVC by VTM-22.0 intra coding. We plot the rate-distortion (RD) curves on Kodak (in-domain) and the other five domains in Fig. \ref{rdcurve}. All pre-trained models perform comparably or better than VTM on Kodak, but most of them fall behind VTM on the other five domains. Our method enhances these models to the level of VTM except on the Pixel domain, where the domain gap is too wide to fully compensate \cite{lv2023dynamic}.

\begin{table*}[ht]
    \centering
    \begin{tabular}{ccccccccc}
        \hline
        $N$ & Baseline & Method & Pixel & Screen-content & Craters & Game & Pathology & AVG \\
        \hline
        \multirow{6}{*}{\;\;5\;\;} & \multirow{3}{*}{Cheng2020} & Finetune & \textbf{-23.09\%} & \textbf{-25.51\%} & 4.35\% & \textbf{-10.69\%} & \textbf{-16.49\%} & -14.29\% \\
                            & ~ & DANICE & -4.92\% & -4.48\% & -7.31\% & -2.65\% & -5.62\% & -5.00\% \\
                           & ~ & Ours & -21.95\% & -23.99\% & \textbf{-8.55\%} & -8.93\% & -13.58\% & \textbf{-15.40\%} \\
                           \cline{2-9}
                           & \multirow{3}{*}{ELIC} & Finetune & -15.82\% & -15.82\% & -3.26\% & -2.73\% & -9.57\% & -9.44\% \\
                           & ~ & DANICE & / & / & / & / & / & / \\
                           & ~ & Ours & \textbf{-17.36\%} & \textbf{-16.27\%} & \textbf{-6.07\%} & \textbf{-4.78\%} & \textbf{-10.70\%} & \textbf{-11.04\%} \\
        \hline
        \multirow{6}{*}{\;\;10\;\;} & \multirow{3}{*}{Cheng2020} & Finetune & \textbf{-27.76\%} & \textbf{-28.42\%} & \textbf{-16.79\%} & \textbf{-13.48\%} & -17.48\% & \textbf{-20.79\%} \\
                            & ~ & DANICE & -8.19\% & -7.78\% & -10.88\% & -4.50\% & -7.14\% & -7.70\% \\
                            & ~ & Ours & -27.44\% & -28.00\% & -15.25\% & -11.34\% & \textbf{-17.78\%} & -19.96\% \\
                            \cline{2-9}
                            & \multirow{3}{*}{ELIC} & Finetune & -20.84\% & -21.92\% & \textbf{-13.06\%} & \textbf{-9.25\%} & -10.41\% & \textbf{-15.10\%} \\
                            & ~ & DANICE & / & / & / & / & / & / \\
                            & ~ & Ours & \textbf{-21.38\%} & \textbf{-22.35\%} & -10.64\% & -6.32\% & \textbf{-11.87\%} & -14.51\% \\
        \hline
        \multirow{12}{*}{\;\;25\;\;} & \multirow{3}{*}{Cheng2020} & Finetune & \textbf{-30.37\%} & \textbf{-34.99\%} & \textbf{-22.83\%} & \textbf{-19.66\%} & -17.72\% & \textbf{-25.11\%} \\
                            & ~ & DANICE & -11.30\% & -11.59\% & -13.58\% & -6.88\% & -9.83\% & -10.64\% \\
                            & ~ & Ours & -27.85\% & -31.10\% & -19.12\% & -14.20\% & \textbf{-18.52\%} & -22.16\% \\
                            \cline{2-9}
                            & \multirow{3}{*}{ELIC} & Finetune & \textbf{-24.63\%} & \textbf{-25.62\%} & \textbf{-17.15\%} & \textbf{-15.95\%} & -10.76\% & \textbf{-18.82\%} \\
                            & ~ & DANICE & / & / & / & / & / & / \\
                            & ~ & Ours & -23.35\% & -24.95\% & -13.77\% & -13.83\% & \textbf{-12.29\%} & -17.64\% \\
                            \cline{2-9}
                            & \multirow{3}{*}{TCM} & Finetune & \textbf{-28.76\%} & \textbf{-15.53\%} & \textbf{-13.16\%} & -12.03\% & -10.21\% & \textbf{-15.94\%} \\
                            & ~ & DANICE & -4.12\% & -4.81\% & -8.52\% & 2.30\% & -4.34\% & -3.90\% \\
                            & ~ & Ours & -27.82\% & -13.81\% & -8.96\% & \textbf{-14.86\%} & \textbf{-10.71\%} & -15.24\% \\ 
                            \cline{2-9}
                            & \multirow{3}{*}{MLIC++} & Finetune & -12.97\% & -9.78\% & \textbf{-11.82\%} & \textbf{-10.61\%} & -9.35\% & -10.91\% \\ 
                            & ~ & DANICE & -0.61\% & -0.79\% & -7.15\% & -0.01\% & -4.90\% & -2.69\% \\
                            & ~ & Ours & \textbf{-14.63\%} & \textbf{-12.09\%} & -8.50\% & -9.70\% & \textbf{-10.27\%} & \textbf{-11.04\%} \\
        \hline
    \end{tabular}
    \caption{Detailed BD-rate ($\downarrow$) using different numbers of target samples ($N$) for domain adaptation. We compare our method with full-model finetune and DANICE. The anchors are corresponding pre-trained models. Note that DANICE can not be applied to ELIC as GDN is not used. Please refer to the \textbf{Appendix B} for more results when $N=\left\{5, 10\right\}$.}
    \label{tb2}
\end{table*}

\begin{table*}[t]
    \centering
    \begin{tabular}{cccccccccc}
    \hline
        \multicolumn{2}{c}{Stage 1} & \multicolumn{2}{c}{Stage 2} & \multirow{2}{*}{Pixel} & \multirow{2}{*}{Screen-content} & \multirow{2}{*}{Craters} & \multirow{2}{*}{Game} & \multirow{2}{*}{Pathology} & \multirow{2}{*}{AVG}\\ 
        \cline{1-4}
        \multicolumn{1}{c}{Conv} & \multicolumn{1}{c}{LoRA} & \multicolumn{1}{c}{Stack 1234} & \multicolumn{1}{c}{Stack 4} & \multicolumn{1}{c}{} & \multicolumn{1}{c}{} & \multicolumn{1}{c}{} & \multicolumn{1}{c}{} & \multicolumn{1}{c}{} & \multicolumn{1}{c}{} \\
        \hline
        \checkmark & & & & -19.38\% & -23.73\% & -10.84\% & -12.73\% & -10.30\% & -15.40\% \\
        \checkmark & \checkmark & & & -20.40\% & -24.28\% & -12.57\% & -13.40\% & -11.50\% & -16.43\% \\
        \checkmark & \checkmark & \checkmark & & -19.92\% & -22.46\% & -11.95\% & -10.68\% & -10.95\% & -15.19\% \\
        \checkmark & \checkmark & & \checkmark & -\textbf{23.35\%} & \textbf{-24.95\%} & \textbf{-13.77\%} & \textbf{-13.83\%} & \textbf{-12.29\%} & \textbf{-17.64\%} \\
        \hline
    \end{tabular}
    \caption{Ablation studies conducted on ELIC adapted with 25 samples. The anchor of BD-rate ($\downarrow$) is pre-trained ELIC models.}
    \label{tb3}
\end{table*}

\subsubsection{Comparisons with Domain Adaptation Method}

To evaluate the efficiency, we also compare with existing DA method DANICE \cite{katakol2021danice} and full-model finetune by RD performance and complexity. We demonstrate detailed adaptation results under different $N$ by computing BD-rate \cite{bjontegaard2001calculation} in Table \ref{tb2}. It is noted that DANICE is restricted to GDN structures \cite{DBLP:journals/corr/BalleLS15}, making it inapplicable to baselines like ELIC. Additionally, DANICE fails to achieve domain adaptation in some conditions, such as TCM and MLIC++ on the Game domain, indicating a limited versatility. Our method, however, achieves excellent adaptation performance in all conditions, performing comparably with full-model finetune and significantly outperforming DANICE. 

In the regard of complexity, we highlight the compactness of adapters in Table \ref{tb1}. The proposed method only needs to transmit fewer than $2\%$ of the total parameters, while much more parameters need considering in DANICE and finetune due to the complex entropy model. Besides, the proposed method costs only a tiny increase on the FLOPs and the decoding time, while enjoys a faster training speed than finetune as suggested in \textbf{Appendix C}.

\subsection{Ablation Studies}

We conduct ablation studies to evaluate the contributions of Conv-Adapters, LoRA-Adapters and the two-stage training strategy. Using pre-trained ELIC as the baseline, we perform the proposed adaptation with 25 samples on the five domains. The results are shown in Table \ref{tb3}. Conv-Adapters achieve an average reduction of 15.40\% in BD-rate compared to pre-trained models, while LoRA-Adapters and the two-stage training strategy each contributes an additional 1\% BD-rate reduction. We also compare our second stage strategy with finetuning all Conv-Adapters in the decoder and display the superiority of our method. 

Given that Conv-Adapter, which involves linear re-allocation on channels, has the most significant impact on adaptation performance, we carry out further explorations on its structure and deployment in \textbf{Appendix D}, and demonstrate the efficiency of the proposed method on both adaptation performance and additional parameters.

\begin{figure*}[htbp]
  \includegraphics[width=\textwidth]{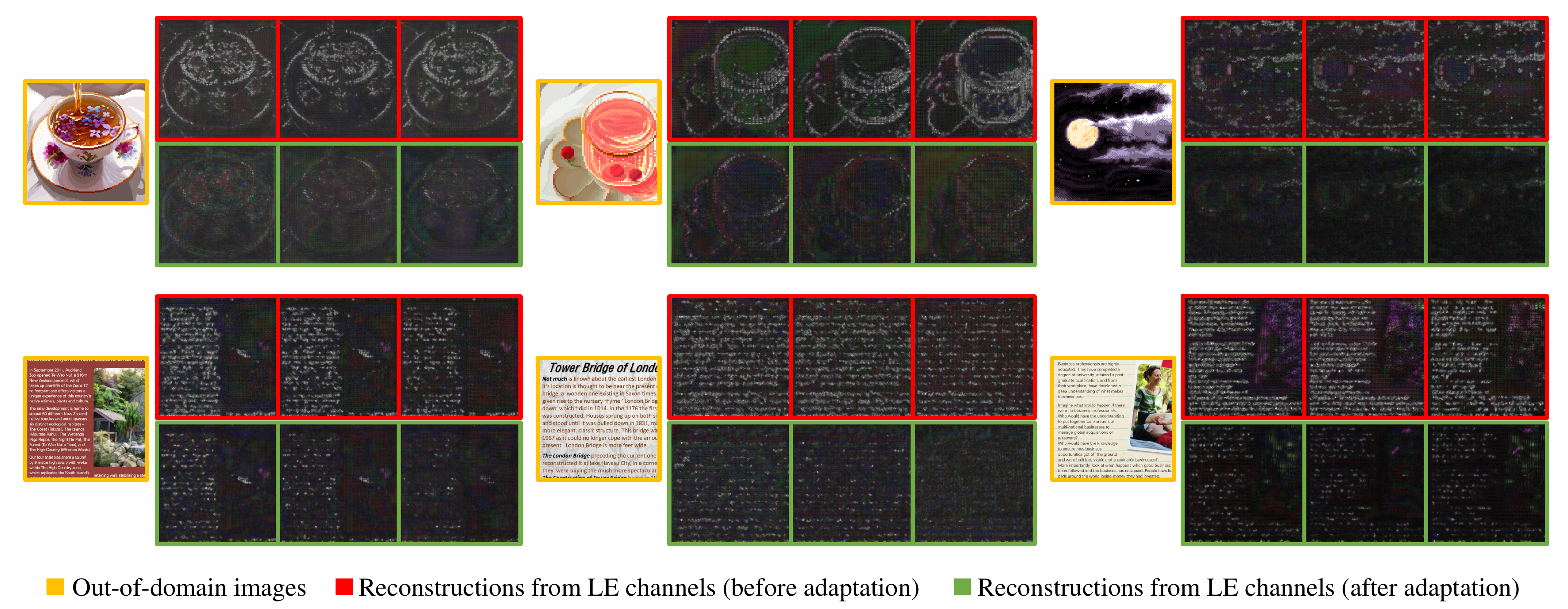}
  \caption{Following Fig. \ref{fig2}, we display reconstructions from low-energy (LE) channels on out-of-domain images. The top row is Pixel, and the bottom row is Screen-content. As observed, our method efficiently reduces the information from LE channels.}
  \label{recon}
\end{figure*}

\begin{figure}[tbp]
  \includegraphics[width=0.47\textwidth]{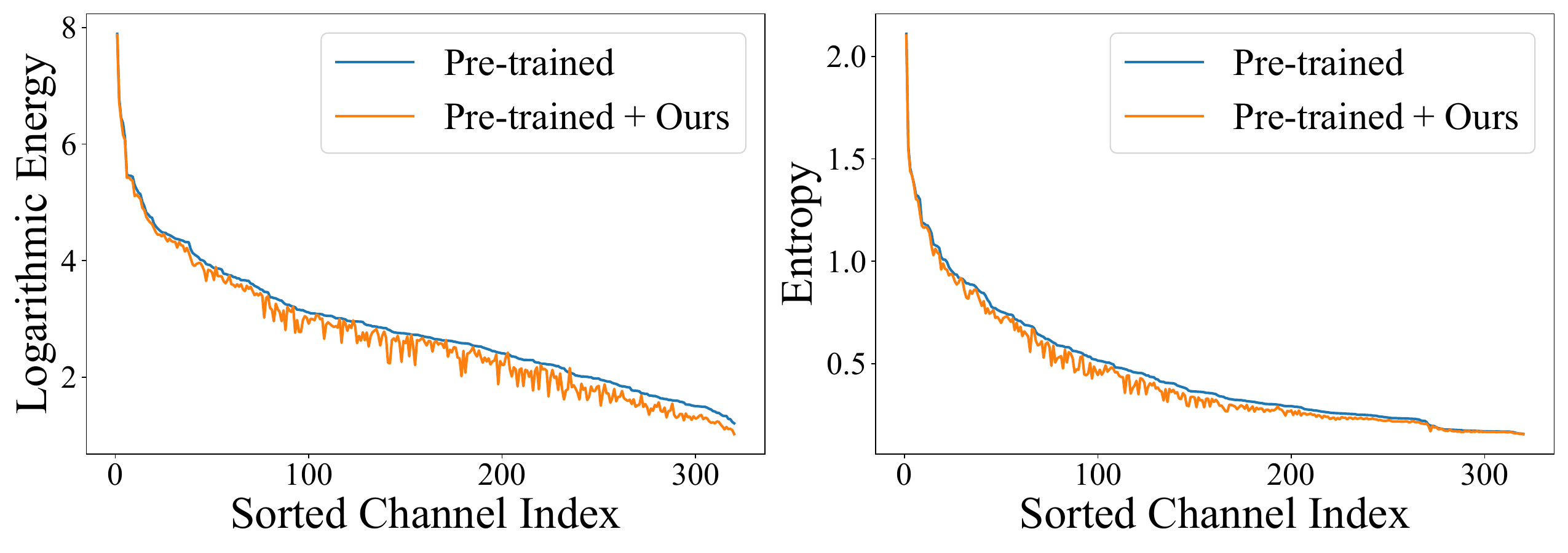}
  \caption{Distributions of energy and entropy on channels. The channels are sorted by the distributions of pre-trained models. The results are averaged on the Pixel domain.}
  \label{channel}
\end{figure}

\subsection{Analyses on Channel-wise Re-allocation}

We propose channel-wise re-allocation with adapters in the pre-trained LIC model for domain adaptation. We get motivated by the observation that the pre-trained channel-wise decomposition is disturbed by out-of-domain images, resulting in scattered energy allocation across channels. To highlight this issue and demonstrate the effectiveness of our method, we perform detailed analyses focusing on the channels of latent representation. We conduct experiments on pre-trained ELIC models. We perform the proposed method on the Pixel domain using 25 target samples.

\subsubsection{Variation on Channels}

We use the output latent of the analysis transform, and study the distribution of logarithmic energy and entropy on channels following \cite{he2022elic} in Fig. \ref{channel}. The results are averaged on the Pixel domain. We find that both energy and entropy on specific channels drop after adapting with adapters. Especially, the variation is more apparent on low-energy channels, while high-energy channels remain almost unchanged. This suggests that the energy becomes more concentrated to high-energy channels on out-of-domain images.

\subsubsection{Reconstructions from Low-Energy Channels}

To further investigate the variation on channel-wise decomposition, we compare the reconstructions from low-energy channels before and after adaptation. Concretely, we keep specific channels unchanged and mask the rest into zeros, reconstruct the modified latent, then subtract the reconstruction of all-zero one. We sort the channels by energy, and display the reconstructions from specific ranges of channels after $200 th$ in Fig. \ref{recon}, which is similar to Fig. \ref{fig2}. We find weaker details of inputs can be observed in reconstructions from the model adapted with our method, suggesting high-frequency components have been better moved out of the low-energy channels, which is more in line with the compact energy allocation in processing in-domain images. Thus, we suggest the pre-trained frequency decomposition on high-frequency components is strengthened by our method. More reconstructions of images from different domains are shown in \textbf{Appendix E}.

\section{Conclusion}

In this paper, we propose a universal few-shot domain adaptation method for learned image compression. With pleasant RD improvement and limited complexity, we expand the proposed method to multiple domains and multiple representative  LIC schemes, demonstrating its versatility. We explore the domain gaps in LIC through disturbed channel-wise decomposition, and introduce compact adapters for channel-wise re-allocation. With our method the domain adaptation of LIC models can be simplified into lightweight training, transmission, and storage of compact adapters, achieving flexibility and efficiency. We conduct sufficient experiments showing the proposed method enhances pre-trained models to the level of VTM generally, and significantly outperforms existing domain adaptation method. We hope our findings provide inspiration for future research on domain adaptation for LIC, thus contributing to the real-world deployment of versatile learned image compression.

\section*{Acknowledgments}

This work was supported by the Natural Science Foundation of China under Grant 62036005. We acknowledge the support of GPU cluster built by MCC Lab of Information Science and Technology Institution, USTC.

\bibliography{aaai25}

\begin{thebibliography}{36}
\providecommand{\natexlab}[1]{#1}

\bibitem[{Agustsson and Timofte(2017)}]{agustsson2017ntire}
Agustsson, E.; and Timofte, R. 2017.
\newblock Ntire 2017 challenge on single image super-resolution: Dataset and study.
\newblock In \emph{Proceedings of the IEEE conference on computer vision and pattern recognition workshops}, 126--135.

\bibitem[{Asuni and Giachetti(2014)}]{asuni2014testimages}
Asuni, N.; and Giachetti, A. 2014.
\newblock TESTIMAGES: a Large-scale Archive for Testing Visual Devices and Basic Image Processing Algorithms.
\newblock In \emph{STAG: Smart Tools \& Apps for Graphics}, 63--70.

\bibitem[{Ball{\'e}, Laparra, and Simoncelli(2015)}]{DBLP:journals/corr/BalleLS15}
Ball{\'e}, J.; Laparra, V.; and Simoncelli, E.~P. 2015.
\newblock Density modeling of images using a generalized normalization transformation.
\newblock \emph{arXiv preprint arXiv:1511.06281}.

\bibitem[{Ball{\'e}, Laparra, and Simoncelli(2016)}]{DBLP:conf/iclr/BalleLS17}
Ball{\'e}, J.; Laparra, V.; and Simoncelli, E.~P. 2016.
\newblock End-to-end optimized image compression.
\newblock \emph{arXiv preprint arXiv:1611.01704}.

\bibitem[{Ball{\'e} et~al.(2018)Ball{\'e}, Minnen, Singh, Hwang, and Johnston}]{DBLP:conf/iclr/BalleMSHJ18}
Ball{\'e}, J.; Minnen, D.; Singh, S.; Hwang, S.~J.; and Johnston, N. 2018.
\newblock Variational image compression with a scale hyperprior.
\newblock \emph{arXiv preprint arXiv:1802.01436}.

\bibitem[{Barman et~al.(2018)Barman, Zadtootaghaj, Schmidt, Martini, and M{\"o}ller}]{barman2018gamingvideoset}
Barman, N.; Zadtootaghaj, S.; Schmidt, S.; Martini, M.~G.; and M{\"o}ller, S. 2018.
\newblock GamingVideoSET: a dataset for gaming video streaming applications.
\newblock In \emph{2018 16th Annual Workshop on Network and Systems Support for Games (NetGames)}, 1--6. IEEE.

\bibitem[{Bjontegaard(2001)}]{bjontegaard2001calculation}
Bjontegaard, G. 2001.
\newblock Calculation of average PSNR differences between RD-curves.
\newblock \emph{ITU SG16 Doc. VCEG-M33}.

\bibitem[{Brancati et~al.(2022)Brancati, Anniciello, Pati, Riccio, Scognamiglio, Jaume, De~Pietro, Di~Bonito, Foncubierta, Botti et~al.}]{brancati2022bracs}
Brancati, N.; Anniciello, A.~M.; Pati, P.; Riccio, D.; Scognamiglio, G.; Jaume, G.; De~Pietro, G.; Di~Bonito, M.; Foncubierta, A.; Botti, G.; et~al. 2022.
\newblock BRACS: A dataset for breast carcinoma subtyping in h\&e histology images.
\newblock \emph{Database}, 2022: baac093.

\bibitem[{Bross et~al.(2021)Bross, Wang, Ye, Liu, Chen, Sullivan, and Ohm}]{bross2021overview}
Bross, B.; Wang, Y.-K.; Ye, Y.; Liu, S.; Chen, J.; Sullivan, G.~J.; and Ohm, J.-R. 2021.
\newblock Overview of the versatile video coding (VVC) standard and its applications.
\newblock \emph{IEEE Transactions on Circuits and Systems for Video Technology}, 31(10): 3736--3764.

\bibitem[{Campos et~al.(2019)Campos, Meierhans, Djelouah, and Schroers}]{DBLP:conf/cvpr/CamposMDS19}
Campos, J.; Meierhans, S.; Djelouah, A.; and Schroers, C. 2019.
\newblock Content adaptive optimization for neural image compression.
\newblock \emph{arXiv preprint arXiv:1906.01223}.

\bibitem[{Cheng et~al.(2020)Cheng, Sun, Takeuchi, and Katto}]{cheng2020learned}
Cheng, Z.; Sun, H.; Takeuchi, M.; and Katto, J. 2020.
\newblock Learned image compression with discretized gaussian mixture likelihoods and attention modules.
\newblock In \emph{Proceedings of the IEEE/CVF conference on computer vision and pattern recognition}, 7939--7948.

\bibitem[{CLIC(2020)}]{clic}
CLIC. 2020.
\newblock Workshop and challenge on learned image compression.
\newblock \url{https://www.compression.cc/}.

\bibitem[{Duan et~al.(2022)Duan, Lu, Ma, and Zhu}]{duan2022opening}
Duan, Z.; Lu, M.; Ma, Z.; and Zhu, F. 2022.
\newblock Opening the black box of learned image coders.
\newblock In \emph{2022 Picture Coding Symposium (PCS)}, 73--77. IEEE.

\bibitem[{Franzen(1993)}]{kodak}
Franzen, R. 1993.
\newblock Kodak lossless true color image suite (photocd pcd0992).
\newblock \url{http://r0k.us/graphics/kodak/}.

\bibitem[{Guo et~al.(2021)Guo, Zhang, Feng, and Chen}]{guo2021soft}
Guo, Z.; Zhang, Z.; Feng, R.; and Chen, Z. 2021.
\newblock Soft then hard: Rethinking the quantization in neural image compression.
\newblock In \emph{International Conference on Machine Learning}, 3920--3929. PMLR.

\bibitem[{He et~al.(2022)He, Yang, Peng, Ma, Qin, and Wang}]{he2022elic}
He, D.; Yang, Z.; Peng, W.; Ma, R.; Qin, H.; and Wang, Y. 2022.
\newblock ELIC: Efficient learned image compression with unevenly grouped space-channel contextual adaptive coding.
\newblock In \emph{Proceedings of the IEEE/CVF Conference on Computer Vision and Pattern Recognition}, 5718--5727.

\bibitem[{Hu et~al.(2021)Hu, Shen, Wallis, Allen-Zhu, Li, Wang, Wang, and Chen}]{DBLP:conf/iclr/HuSWALWWC22}
Hu, E.~J.; Shen, Y.; Wallis, P.; Allen-Zhu, Z.; Li, Y.; Wang, S.; Wang, L.; and Chen, W. 2021.
\newblock LoRA: Low-rank adaptation of large language models.
\newblock \emph{arXiv preprint arXiv:2106.09685}.

\bibitem[{Jiang(2022)}]{jiang2022unofficialelic}
Jiang, W. 2022.
\newblock Unofficial ELIC.
\newblock \url{https://github.com/JiangWeibeta/ELIC}.

\bibitem[{Jiang and Wang(2023)}]{jiang2023mlicpp}
Jiang, W.; and Wang, R. 2023.
\newblock MLIC++: Linear Complexity Multi-Reference Entropy Modeling for Learned Image Compression.
\newblock In \emph{ICML 2023 Workshop Neural Compression: From Information Theory to Applications}.

\bibitem[{Jiang et~al.(2023)Jiang, Yang, Zhai, Ning, Gao, and Wang}]{jiang2023mlic}
Jiang, W.; Yang, J.; Zhai, Y.; Ning, P.; Gao, F.; and Wang, R. 2023.
\newblock MLIC: Multi-reference entropy model for learned image compression.
\newblock In \emph{Proceedings of the 31st ACM International Conference on Multimedia}, 7618--7627.

\bibitem[{Katakol et~al.(2021)Katakol, Herranz, Yang, and Mrak}]{katakol2021danice}
Katakol, S.; Herranz, L.; Yang, F.; and Mrak, M. 2021.
\newblock DANICE: Domain adaptation without forgetting in neural image compression.
\newblock In \emph{Proceedings of the IEEE/CVF Conference on Computer Vision and Pattern Recognition}, 1921--1925.

\bibitem[{Li et~al.(2024{\natexlab{a}})Li, Dai, Fang, Zheng, Fei, Xiong, and Zhang}]{li2023revisiting}
Li, S.; Dai, W.; Fang, Y.; Zheng, Z.; Fei, W.; Xiong, H.; and Zhang, W. 2024{\natexlab{a}}.
\newblock Revisiting Learned Image Compression With Statistical Measurement of Latent Representations.
\newblock \emph{IEEE Transactions on Circuits and Systems for Video Technology}, 34(4): 2891--2907.

\bibitem[{Li et~al.(2024{\natexlab{b}})Li, Liao, Tang, Zhang, Li, Bian, Sheng, Feng, Li, Gao et~al.}]{li2024ustc}
Li, Z.; Liao, J.; Tang, C.; Zhang, H.; Li, Y.; Bian, Y.; Sheng, X.; Feng, X.; Li, Y.; Gao, C.; et~al. 2024{\natexlab{b}}.
\newblock USTC-TD: A Test Dataset and Benchmark for Image and Video Coding in 2020s.
\newblock \emph{arXiv preprint arXiv:2409.08481}.

\bibitem[{Liu, Sun, and Katto(2023)}]{liu2023learned}
Liu, J.; Sun, H.; and Katto, J. 2023.
\newblock Learned image compression with mixed transformer-cnn architectures.
\newblock In \emph{Proceedings of the IEEE/CVF Conference on Computer Vision and Pattern Recognition}, 14388--14397.

\bibitem[{Lv et~al.(2023)Lv, Xiang, Zhang, Yang, Han, and Yang}]{lv2023dynamic}
Lv, Y.; Xiang, J.; Zhang, J.; Yang, W.; Han, X.; and Yang, W. 2023.
\newblock Dynamic Low-Rank Instance Adaptation for Universal Neural Image Compression.
\newblock In \emph{Proceedings of the 31st ACM International Conference on Multimedia}, 632--642.

\bibitem[{Minnen, Ball{\'e}, and Toderici(2018)}]{minnen2018joint}
Minnen, D.; Ball{\'e}, J.; and Toderici, G.~D. 2018.
\newblock Joint autoregressive and hierarchical priors for learned image compression.
\newblock \emph{Advances in neural information processing systems}, 31: 10794--10803.

\bibitem[{Ni et~al.(2017)Ni, Ma, Zeng, Chen, Cai, and Ma}]{ni2017esim}
Ni, Z.; Ma, L.; Zeng, H.; Chen, J.; Cai, C.; and Ma, K.-K. 2017.
\newblock ESIM: Edge similarity for screen content image quality assessment.
\newblock \emph{IEEE Transactions on Image Processing}, 26(10): 4818--4831.

\bibitem[{Presta et~al.(2024)Presta, Spadaro, Tartaglione, Fiandrotti, and Grangetto}]{presta2024domain}
Presta, A.; Spadaro, G.; Tartaglione, E.; Fiandrotti, A.; and Grangetto, M. 2024.
\newblock Domain Adaptation for Learned Image Compression with Supervised Adapters.
\newblock In \emph{2024 Data Compression Conference (DCC)}, 33--42. IEEE.

\bibitem[{Robinson et~al.(2010)Robinson, Brylow, Tschimmel, Humm, Lawrence, Thomas, Denevi, Bowman-Cisneros, Zerr, Ravine et~al.}]{robinson2010lunar}
Robinson, M.~S.; Brylow, S.; Tschimmel, M.; Humm, D.; Lawrence, S.; Thomas, P.; Denevi, B.~W.; Bowman-Cisneros, E.; Zerr, J.; Ravine, M.; et~al. 2010.
\newblock Lunar reconnaissance orbiter camera (LROC) instrument overview.
\newblock \emph{Space science reviews}, 150: 81--124.

\bibitem[{Shen, Yue, and Yang(2023)}]{shen2023dec}
Shen, S.; Yue, H.; and Yang, J. 2023.
\newblock Dec-adapter: Exploring efficient decoder-side adapter for bridging screen content and natural image compression.
\newblock In \emph{Proceedings of the IEEE/CVF International Conference on Computer Vision}, 12887--12896.

\bibitem[{Sullivan et~al.(2012)Sullivan, Ohm, Han, and Wiegand}]{sullivan2012overview}
Sullivan, G.~J.; Ohm, J.-R.; Han, W.-J.; and Wiegand, T. 2012.
\newblock Overview of the high efficiency video coding (HEVC) standard.
\newblock \emph{IEEE Transactions on circuits and systems for video technology}, 22(12): 1649--1668.

\bibitem[{Tsubota, Akutsu, and Aizawa(2023)}]{tsubota2023universal}
Tsubota, K.; Akutsu, H.; and Aizawa, K. 2023.
\newblock Universal deep image compression via content-adaptive optimization with adapters.
\newblock In \emph{Proceedings of the IEEE/CVF Winter Conference on Applications of Computer Vision}, 2529--2538.

\bibitem[{Van~Rozendaal et~al.(2021)Van~Rozendaal, Brehmer, Zhang, Pourreza, Wiggers, and Cohen}]{vanrozendaal2021instanceadaptive}
Van~Rozendaal, T.; Brehmer, J.; Zhang, Y.; Pourreza, R.; Wiggers, A.; and Cohen, T.~S. 2021.
\newblock Instance-adaptive video compression: Improving neural codecs by training on the test set.
\newblock \emph{arXiv preprint arXiv:2111.10302}.

\bibitem[{Wallace(1991)}]{wallace1991jpeg}
Wallace, G.~K. 1991.
\newblock The JPEG still picture compression standard.
\newblock \emph{Communications of the ACM}, 34(4): 30--44.

\bibitem[{Yang, Fang, and Lin(2015)}]{yang2015perceptual}
Yang, H.; Fang, Y.; and Lin, W. 2015.
\newblock Perceptual quality assessment of screen content images.
\newblock \emph{IEEE Transactions on Image Processing}, 24(11): 4408--4421.

\bibitem[{Yang et~al.(2021)Yang, Shen, Yue, and Li}]{yang2021implicit}
Yang, J.; Shen, S.; Yue, H.; and Li, K. 2021.
\newblock Implicit transformer network for screen content image continuous super-resolution.
\newblock \emph{Advances in Neural Information Processing Systems}, 34: 13304--13315.

\end{thebibliography}

\clearpage
\appendix
\setcounter{secnumdepth}{2}
\section*{Appendix}

\section{Comparisons with Instance Adaptation}
\begin{figure}[t]
\centering
  \includegraphics[width=0.48\textwidth]{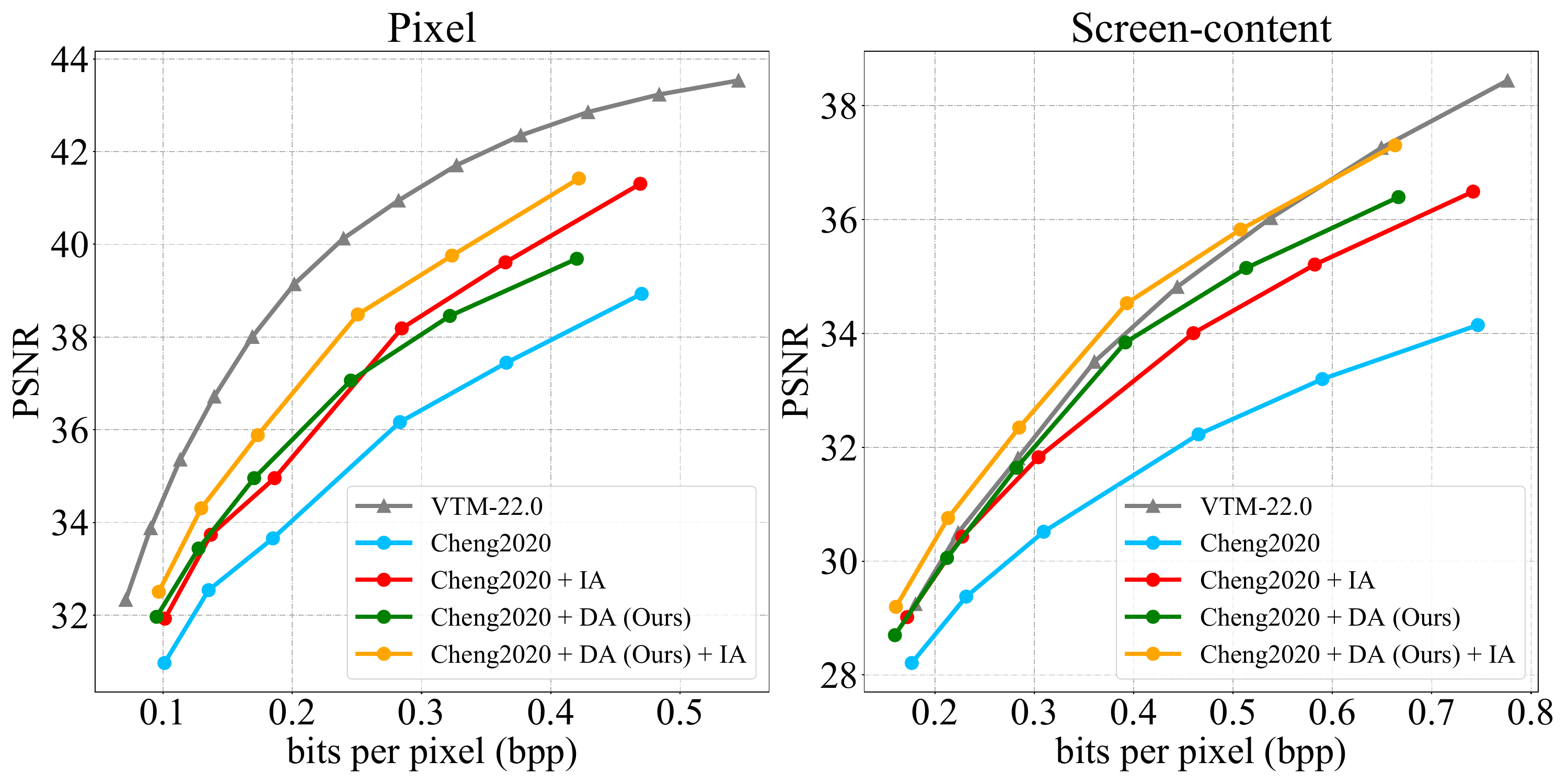}
  \caption{Compatibility between our method and representative instance adaptation method \cite{lv2023dynamic} (abbreviated as DA and IA, respectively).}
  \label{instance}
\end{figure}

In this section, we present our proposed domain adaptation (DA) method for learned image compression (LIC), comparing it to existing instance adaptation (IA) methods \cite{DBLP:conf/cvpr/CamposMDS19, vanrozendaal2021instanceadaptive, tsubota2023universal, shen2023dec, lv2023dynamic}. IA adapts a pre-trained LIC model to individual images through per-image online training during inference. In contrast, DA involves an one-time training process on multiple images, adapting the model to a target domain before inference. Consequently, we suggest that our proposed DA method for LIC complements existing IA methods. The compatibility between these methods is illustrated in Fig. \ref{instance}.

Specifically, we explore the combination of our proposed method with one of the latest IA methods \cite{lv2023dynamic}. We apply both techniques to the Pixel and Screen-content domain. For alignment, we use the pre-trained Cheng2020 model \cite{cheng2020learned} as the baseline. Our DA method is conducted with 25 target samples, followed by implementing IA for each image on the adapted model. The rate-distortion (RD) curves, compared with VTM-22.0 intra coding, the pre-trained Cheng2020, and each technique individually, are shown in Fig. \ref{instance}. Note that some curves differ slightly due to the different padding method adopted by \citet{lv2023dynamic} in their released code; we respect their implementation in Fig. \ref{instance}. When combining both techniques, the RD performance is further improved, indicating that our method complements IA by introducing additional domain information from a limited number of samples.

\begin{figure}[t]
\centering
  \includegraphics[width=0.48\textwidth]{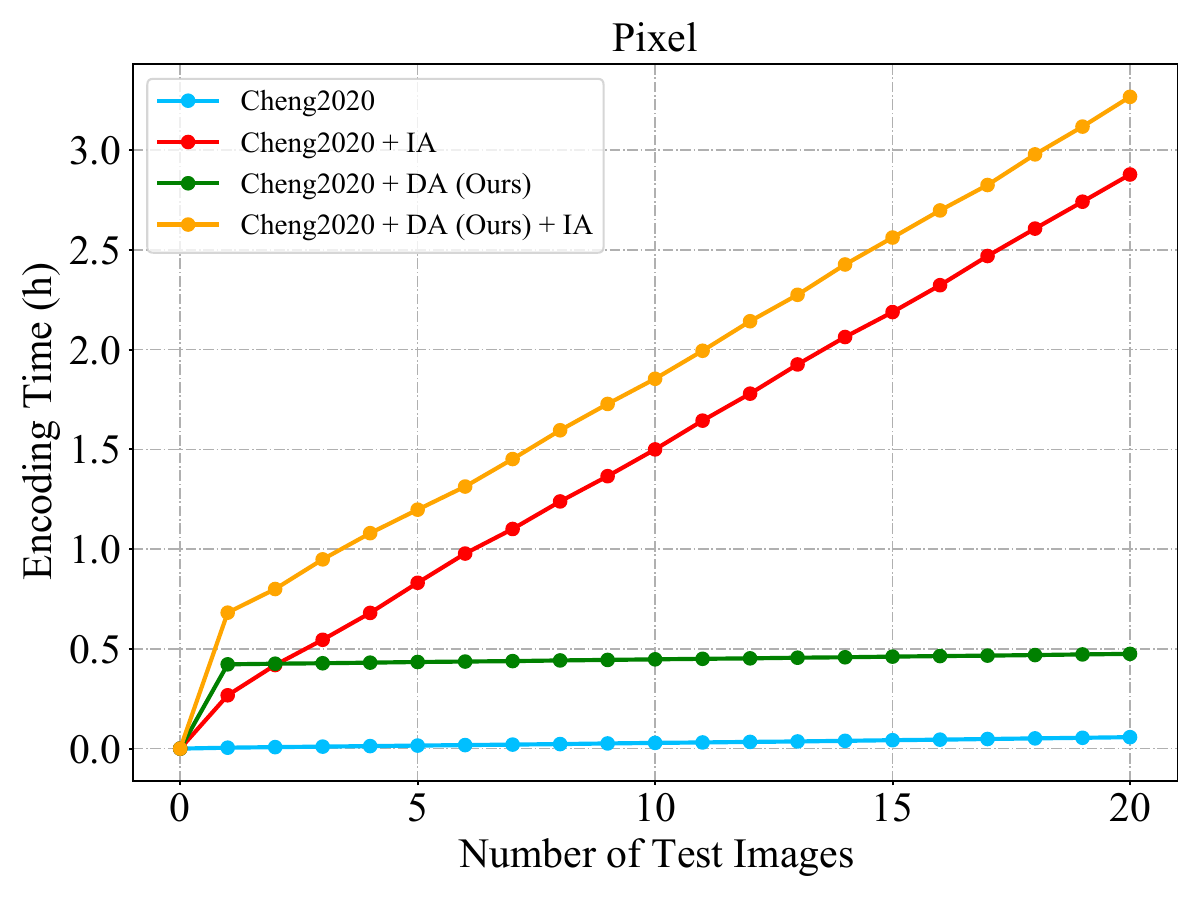}
  \caption{Encoding time of different methods when applied to different numbers of test images from the Pixel domain, using a single Tesla V100 GPU. The proposed few-shot domain adaptation method is conducted on 25 target samples.}
  \label{time}
\end{figure}

For encoding time, IA relies on expensive per-image online training, whereas DA is more efficient due to its one-time training on multiple samples. We display the encoding time for processing different numbers of test images in Fig. \ref{time}. For DA, a fixed one-time training is required to adapt the pre-trained LIC model to the target domain. Adapting the pre-trained Cheng2020 model with 25 samples takes less than half an hour on a single Tesla V100 GPU. Once adapted, the model achieves the same encoding speed as the source model. In contrast, IA requires time-consuming online training for every test image, resulting in a significantly longer encoding time compared to the source model. Therefore, DA is considerably more efficient than IA when encoding a batch of target-domain images.

\section{Detailed Domain Adaptation Performance}

We present the detailed domain adaptation performance in Table \ref{tb6}, which is a superset of Table \ref{tb2}. We compare our proposed method with the existing domain adaptation method DANICE \cite{katakol2021danice}, and full-model finetune using different numbers of target-domain samples. To evaluate the performance, we compute the BD-rate \cite{bjontegaard2001calculation} against the source model. It should be noted that DANICE is restricted to GDN structures \cite{DBLP:journals/corr/BalleLS15}, making it inapplicable to baselines like ELIC \cite{he2022elic}. Our method outperforms DANICE in all conditions and performs comparably with full-model fine-tuning.

\begin{table}[!h]
    \centering
    \setlength{\tabcolsep}{1mm}
    \begin{tabular}{ccccc}
    \hline
        \multirow{2}{*}{Baseline} & \multicolumn{2}{c}{Dec. T (s)} & \multicolumn{2}{c}{MFLOPs / pixel}\\
        \cline{2-5}
        \multicolumn{1}{c}{} & \multicolumn{1}{c}{Source} & \multicolumn{1}{c}{Ours} & \multicolumn{1}{c}{Source} & \multicolumn{1}{c}{Ours} \\
        \hline
        Cheng2020 & $9.570$ & $9.588_{+0.19\%}$ & $1.03$ & $1.08_{+5.21\%}$ \\
        ELIC & $0.319$ & $0.324_{+1.50\%}$ & $0.85$ & $0.92_{+9.22\%}$ \\
        TCM & $0.269$ & $0.270_{+0.60\%}$ & $0.54$ & $0.58_{+7.07\%}$ \\
        MLIC++ & $0.338$ & $0.342_{+1.30\%}$ & $1.26$ & $1.31_{+3.90\%}$ \\
        \hline
    \end{tabular}
    \caption{Complexity compared with source models. The results are averaged on Kodak with a single Tesla V100 GPU.}
    \label{tb4}
\end{table}

\begin{table}[!h]
    \centering
    \begin{tabular}{ccc}
    \hline
        Baseline & Finetune & Ours \\ 
        \hline
        Cheng2020 & 0.246 & 0.216 \\
        ELIC & 0.418 & 0.356 \\
        TCM & 0.672 & 0.536 \\
        MLIC++ & 0.574 & 0.512 \\
        \hline
    \end{tabular}
    \caption{Comparisons of training speed (second per iteration) between our method and full-model finetune.}
    \label{tb5}
\end{table}

\begin{table*}[!ht]
    \centering
    \begin{tabular}{ccccccccc}
        \hline
        $N$ & Baseline & Method & Pixel & Screen-content & Craters & Game & Pathology & AVG \\
        \hline
        \multirow{12}{*}{\;\;5\;\;} & \multirow{3}{*}{Cheng2020} & Finetune & \textbf{-23.09\%} & \textbf{-25.51\%} & 4.35\% & \textbf{-10.69\%} & \textbf{-16.49\%} & -14.29\% \\
                            & ~ & DANICE & -4.92\% & -4.48\% & -7.31\% & -2.65\% & -5.62\% & -5.00\% \\
                           & ~ & Ours & -21.95\% & -23.99\% & \textbf{-8.55\%} & -8.93\% & -13.58\% & \textbf{-15.40\%} \\
                           \cline{2-9}
                           & \multirow{3}{*}{ELIC} & Finetune & -15.82\% & -15.82\% & -3.26\% & -2.73\% & -9.57\% & -9.44\% \\
                           & ~ & DANICE & / & / & / & / & / & / \\
                           & ~ & Ours & \textbf{-17.36\%} & \textbf{-16.27\%} & \textbf{-6.07\%} & \textbf{-4.78\%} & \textbf{-10.70\%} & \textbf{-11.04\%} \\
                           \cline{2-9}
                           & \multirow{3}{*}{TCM} & Finetune & -17.88\% & \textbf{-7.16\%} & \textbf{-4.75\%} & \textbf{-4.78\%} & -7.87\% & -8.49\% \\
                           & ~ & DANICE & -1.66\% & -1.64\% & -3.20\% & -0.85\% & -3.17\% & -2.10\% \\
                           & ~ & Ours & \textbf{-20.10\%} & -6.61\% & -3.31\% & -4.23\% & \textbf{-9.40\%} & \textbf{-8.73\%} \\ 
                           \cline{2-9}
                           & \multirow{3}{*}{MLIC++} & Finetune & -3.23\% & -3.09\% & -1.24\% & \textbf{-3.31\%} & -6.93\% & -3.56\% \\
                           & ~ & DANICE & -0.60\% & -0.33\% & -2.63\% & 0.12\% & -4.49\% & -1.63\% \\
                           & ~ & Ours & \textbf{-9.84\%} & \textbf{-5.41\%} & \textbf{-1.62\%} & -2.46\% & \textbf{-8.13\%} & \textbf{-5.49\%} \\
        \hline
        \multirow{12}{*}{\;\;10\;\;} & \multirow{3}{*}{Cheng2020} & Finetune & \textbf{-27.76\%} & \textbf{-28.42\%} & \textbf{-16.79\%} & \textbf{-13.48\%} & -17.48\% & \textbf{-20.79\%} \\
                            & ~ & DANICE & -8.19\% & -7.78\% & -10.88\% & -4.50\% & -7.14\% & -7.70\% \\
                            & ~ & Ours & -27.44\% & -28.00\% & -15.25\% & -11.34\% & \textbf{-17.78\%} & -19.96\% \\
                            \cline{2-9}
                            & \multirow{3}{*}{ELIC} & Finetune & -20.84\% & -21.92\% & \textbf{-13.06\%} & \textbf{-9.25\%} & -10.41\% & \textbf{-15.10\%} \\
                            & ~ & DANICE & / & / & / & / & / & / \\
                            & ~ & Ours & \textbf{-21.38\%} & \textbf{-22.35\%} & -10.64\% & -6.32\% & \textbf{-11.87\%} & -14.51\% \\
                            \cline{2-9}
                            & \multirow{3}{*}{TCM} & Finetune & -24.06\% & \textbf{-13.08\%} & \textbf{-8.95\%} & -7.28\% & -9.67\% & \textbf{-12.61\%} \\
                            & ~ & DANICE & -2.95\% & -3.11\% & -5.52\% & 1.35\% & -3.67\% & -2.78\% \\
                            & ~ & Ours & \textbf{-24.18\%} & -11.87\% & -7.06\% & \textbf{-9.01\%} & \textbf{-10.24\%} & -12.47\% \\ 
                            \cline{2-9}
                            & \multirow{3}{*}{MLIC++} & Finetune & -8.71\% & -8.93\% & \textbf{-7.84\%} & -4.07\% & -8.90\% & -7.69\% \\
                            & ~ & DANICE & -0.32\% & -0.37\% & -4.22\% & 1.52\% & -4.97\% & -1.67\% \\
                            & ~ & Ours & \textbf{-13.82\%} & \textbf{-9.95\%} & -5.66\% & \textbf{-5.32\%} & \textbf{-9.16\%} & \textbf{-8.78\%} \\
        \hline
        \multirow{12}{*}{\;\;25\;\;} & \multirow{3}{*}{Cheng2020} & Finetune & \textbf{-30.37\%} & \textbf{-34.99\%} & \textbf{-22.83\%} & \textbf{-19.66\%} & -17.72\% & \textbf{-25.11\%} \\
                            & ~ & DANICE & -11.30\% & -11.59\% & -13.58\% & -6.88\% & -9.83\% & -10.64\% \\
                            & ~ & Ours & -27.85\% & -31.10\% & -19.12\% & -14.20\% & \textbf{-18.52\%} & -22.16\% \\
                            \cline{2-9}
                            & \multirow{3}{*}{ELIC} & Finetune & \textbf{-24.63\%} & \textbf{-25.62\%} & \textbf{-17.15\%} & \textbf{-15.95\%} & -10.76\% & \textbf{-18.82\%} \\
                            & ~ & DANICE & / & / & / & / & / & / \\
                            & ~ & Ours & -23.35\% & -24.95\% & -13.77\% & -13.83\% & \textbf{-12.29\%} & -17.64\% \\
                            \cline{2-9}
                            & \multirow{3}{*}{TCM} & Finetune & \textbf{-28.76\%} & \textbf{-15.53\%} & \textbf{-13.16\%} & -12.03\% & -10.21\% & \textbf{-15.94\%} \\
                            & ~ & DANICE & -4.12\% & -4.81\% & -8.52\% & 2.30\% & -4.34\% & -3.90\% \\
                            & ~ & Ours & -27.82\% & -13.81\% & -8.96\% & \textbf{-14.86\%} & \textbf{-10.71\%} & -15.24\% \\ 
                            \cline{2-9}
                            & \multirow{3}{*}{MLIC++} & Finetune & -12.97\% & -9.78\% & \textbf{-11.82\%} & \textbf{-10.61\%} & -9.35\% & -10.91\% \\ 
                            & ~ & DANICE & -0.61\% & -0.79\% & -7.15\% & -0.01\% & -4.90\% & -2.69\% \\
                            & ~ & Ours & \textbf{-14.63\%} & \textbf{-12.09\%} & -8.50\% & -9.70\% & \textbf{-10.27\%} & \textbf{-11.04\%} \\
        \hline
    \end{tabular}
    \caption{Detailed BD-rate ($\downarrow$) using different numbers of target samples ($N$) for domain adaptation.}
    \label{tb6}
\end{table*}

\section{Computational Complexity}

In this section, we examine the variations in decoding time and FLOPs introduced by the proposed adapters. The results are presented in Table \ref{tb4}. Compared to the source models, our adapters increase FLOPs by less than 10\% and coding time by less than 1.5\%. Additionally, we provide the training speed for the four baselines we used, mainly comparing our method with full-model fine-tuning. Using $256 \times 256$ patches as input and a batch size of 4, the results are shown in Table \ref{tb5}, obtained with a single Tesla V100 GPU. The training speeds of both methods are of the same magnitude, with our method being slightly faster.

\section{Explorations on Structure and Deployment of Conv-Adapters}

We further explore the structure of Conv-Adapters in Table \ref{tb7} and the deployment strategy in Table \ref{tb8}. We conduct adaptation on ELIC using 25 target samples and evaluate the performance by parameters and rate-distortion improvement.

\begin{table*}[!ht]
    \centering
    \setlength{\tabcolsep}{1mm}
    \begin{tabular}{cccccccc}
    \hline
        Structure & Pixel & Screen-content & Craters & Game & Pathology & AVG & Transmit \\ 
        \hline
        GDN & -7.69\% & -9.93\% & -7.85\% & -3.38\% & -7.99\% & -7.37\% & 0.26M \\
        Depthwise Conv3$\times$3 & -11.31\% & -15.41\% & -9.98\% & -5.42\% & -11.06\% & -10.63\% & 0.26M \\
        Depthwise Conv3$\times$3 + Conv1$\times$1 & -20.86\% & -24.02\% & -13.15\% & -11.40\% & \textbf{-12.53}\% & -16.39\% & 0.73M \\
        Conv3$\times$3 & -22.49\% & -24.44\% & -13.32\% & \textbf{-15.11}\% & -11.60\% & -17.39\% & 4.48M \\
        Conv1$\times$1 (Ours) & -\textbf{23.35\%} & \textbf{-24.95\%} & \textbf{-13.77\%} & -13.83\% & -12.29\% & \textbf{-17.64\%} & 0.71M \\
        \hline
    \end{tabular}
    \caption{Explorations on the structures of Conv-Adapters. Evaluated by the parameters to transmit and the BD-rate ($\downarrow$).}
    \label{tb7}
\end{table*}

\begin{table*}[!ht]
    \centering
    \setlength{\tabcolsep}{1mm}
    \begin{tabular}{ccccccccccc}
    \hline
        Stack 1 & Stack 2 & Stack 3 & Stack 4 & Pixel & Screen-content & Craters & Game & Pathology & AVG & Transmit \\ 
        \hline
        \checkmark & & & & -4.89\% & -4.89\% & -6.64\% & -2.65\% & -8.24\% & -5.46\% & 0.34M \\
        \checkmark & \checkmark & & & -6.70\% & -7.44\% & -7.83\% & -3.81\% & -10.23\% & -7.20\% & 0.45M \\
        \checkmark & \checkmark & \checkmark & & -12.23\% & -14.48\% & -10.64\% & -6.96\% & 13.48\% & -11.56\% & 0.60M \\
        \checkmark & \checkmark & \checkmark & \checkmark & -\textbf{23.35\%} & \textbf{-24.95\%} & \textbf{-13.77\%} & \textbf{-13.83\%} & \textbf{-12.29\%} & \textbf{-17.64\%} & 0.71M \\
        \hline
    \end{tabular}
    \caption{Explorations on the deployment of Conv-Adapters. Evaluated by the parameters to transmit and the BD-rate ($\downarrow$).}
    \label{tb8}
\end{table*}

For structure, we first replace the proposed Conv1$\times$1 with more compact structures such as GDN \cite{DBLP:journals/corr/BalleLS15} and depthwise Conv3$\times$3. We then test more complex structures, including Conv3$\times$3 and a sequence of depthwise Conv3$\times$3 followed by Conv1$\times$1. For all structures, we follow the restriction on initialization and initialize them to perform identical transforms at the start of training. Using ELIC \cite{he2022elic} as the baseline model, we conduct adaptation experiments on five domains, each with 25 target samples. The training strategy for adapters remains consistent. As shown in Table \ref{tb7}, compact structures like GDN and depthwise Conv3$\times$3 have fewer parameters to transmit but exhibit significant degradation in average adaptation performance. For more complex structures, the increased parameters in the spatial dimension fail to further improve performance, underscoring the efficiency of Conv1$\times$1 in balancing complexity and performance.

\begin{figure}[ht]
\centering
  \includegraphics[width=0.48\textwidth]{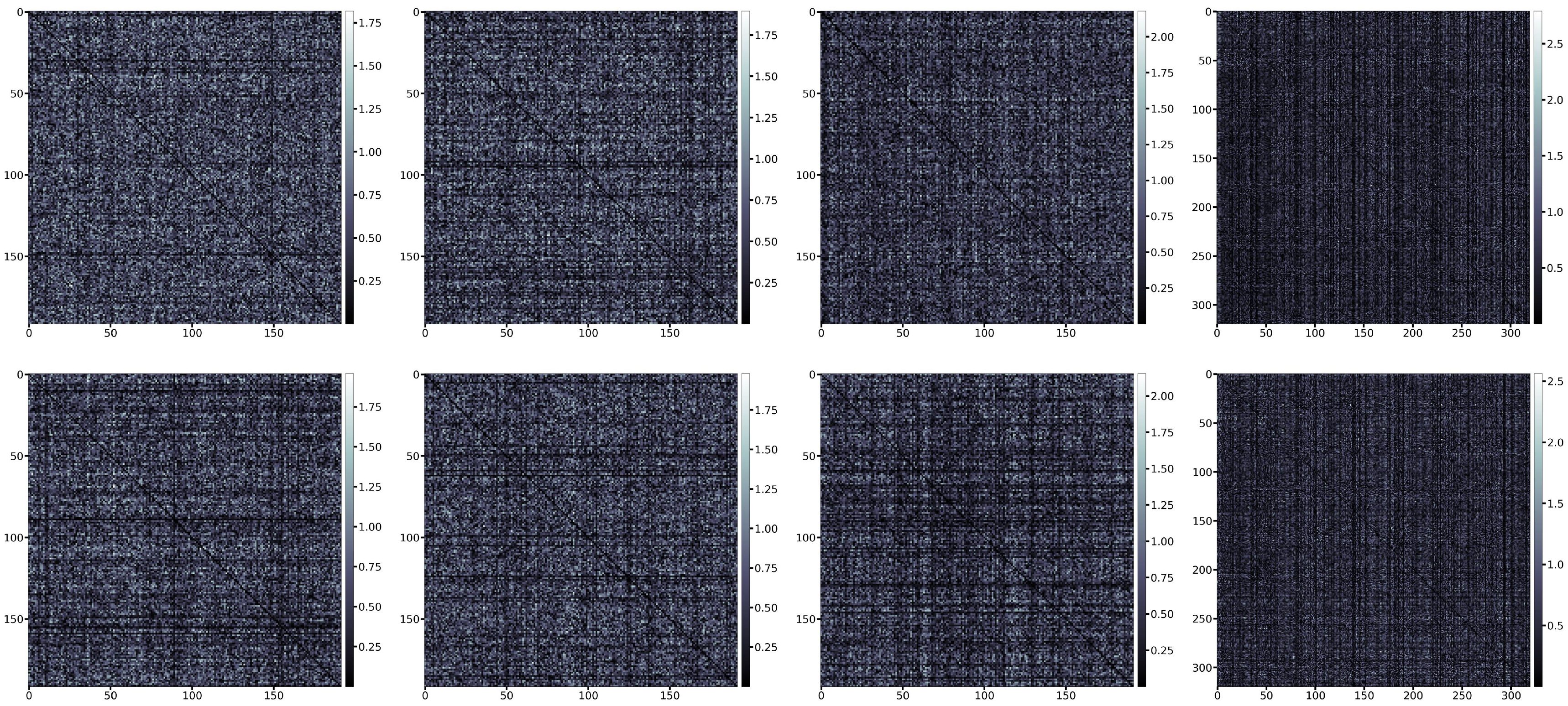}
  \caption{Visualization of Conv-Adapter kernels after training. Four pairs of kernels symmetrical in transform are displayed in two rows, where the top row comes from $g_{a}$ and the bottom row comes from $g_{s}$. We find obvious relevance on rows and columns of the kernels, suggesting channel-wise correlations remain in the latents have been captured.}
  \label{kernel}
\end{figure}

For deployment, following Fig. 5, we divide the analysis transform and the synthesis transform into four stacks and gradually apply Conv-Adapters to each stack. Although the parameters to transmit increase, applying Conv-Adapters to all stacks significantly enhances adaptation performance. Therefore, we recommend deploying Conv-Adapters in all stacks. We visualize the kernels after training in Fig. \ref{kernel}.

\section{Visualization of Reconstructions}

We present the reconstructions of the pre-trained ELIC \cite{he2022elic}, the adapted ELIC using 25 samples, and the corresponding reconstruction errors in Fig. \ref{sm1}. The model is optimized using MSE with $\lambda=0.483$. We showcase out-of-domain samples from the five domains we used: Pixel, Screen-content, Craters, Games, and Pathology. Specific $50 \times 50$ crops from the reconstructions are displayed with pixel enlargement for better visibility. In addition to RD improvement, reconstructions from the adapted model demonstrate superior visual quality, particularly in high-frequency edges and details. For Pixel and Screen-content samples, the sharp edges of pixel-style art and letters are reconstructed more accurately. For samples from Craters, Games, and Pathology, our reconstructions capture more details of the ground truth. We also compute the reconstruction errors relative to the ground truth, suggesting our method primarily reduces reconstruction errors at edges and in fine details.

\begin{figure}[!ht]
\centering
  \includegraphics[width=0.45\textwidth]{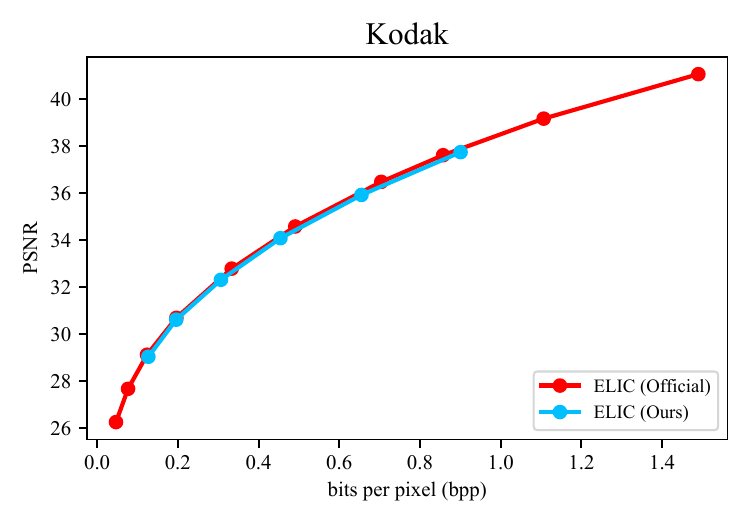}
  \caption{Performance compared with official results.}
  \label{elic}
\end{figure}

\section{Details for Reproducing ELIC}

We utilize DIV2K \cite{agustsson2017ntire} and CLIC 2020 Professional \cite{clic} for training ELIC \cite{he2022elic} models. We construct our code on the base of \cite{jiang2022unofficialelic}. During training, images are randomly cropped into 256 × 256 patches, and a batch size of 8 is adopted following official implementation. We optimize the model using distortion terms of MSE and set the Lagrange multiplier $\lambda \in \left\{18, 35, 67, 130, 250, 483\right\} \times 10^{-4}$ for different quality presets. We use Adam optimizer with $\beta _{1}=0.9$, $\beta _{2}=0.999$. We train each model for 2M steps. The learning rate starts from $10^{-4}$ and drops to $3\times 10^{-5}$ at 1.2M steps, then to $10^{-5}$ at 1.5M steps, and finally to $10^{-6}$ at 1.8M steps. The performance of our models is generally consistent with the official-released results as shown in Fig. \ref{elic}. It is noted that a different set of $\lambda$ was adopted in official results, where the 6 median values were $\left\{16, 32, 75, 150, 300, 450\right\} \times 10^{-4}$.

\begin{figure*}[ht]
\centering
  \includegraphics[width=\textwidth]{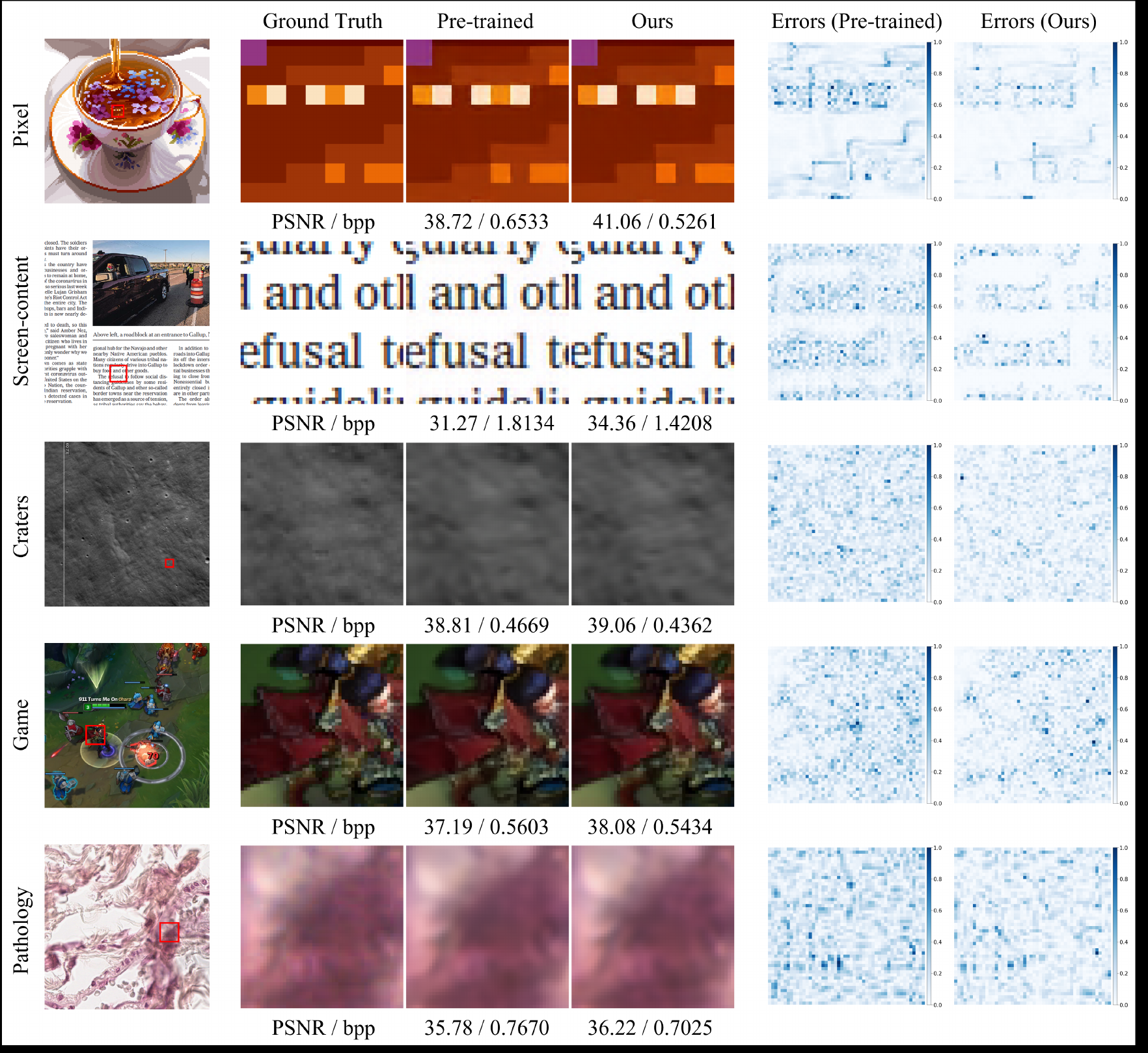}
  \caption{Reconstructions and PSNR ($\uparrow$) / bpp ($\downarrow$) comparisons, detailed by specific 50 × 50 crops. The reconstruction errors with ground truth are shown in right.}
  \label{sm1}
\end{figure*}

\end{document}